\documentclass{aa}
\usepackage[T1]{fontenc}
\usepackage[utf8]{inputenc}

\usepackage{newtxtext}  
\usepackage[varg]{newtxmath}  
\usepackage{amsmath}    
\usepackage{amssymb}    
\usepackage{marvosym}   
\usepackage{nicefrac}   
\usepackage{textcomp}
\usepackage{url}
\usepackage{bm}         
\usepackage{mathtools}
\usepackage{microtype}
\usepackage{multirow}

\usepackage[exponent-product=\cdot
            ,tight-spacing=true
            ,range-phrase={\text{--}}
            ,range-units=single
            ,list-units=single
            ,angle-symbol-over-decimal=true
            ,detect-weight=true
            ,list-pair-separator= {\text{ to }}
            ,list-final-separator= {\text{, and }}
            ]{siunitx}
\DeclareSIUnit\au{au}
\DeclareSIUnit\pc{pc}
\DeclareSIUnit\jy{Jy}
\DeclareSIUnit\msun{M\ensuremath{_{\sun}}}
\DeclareSIUnit\lsun{L\ensuremath{_{\sun}}}

\usepackage{graphicx}
    \DeclareGraphicsExtensions{.pdf}
    \setkeys{Gin}{width=\linewidth}
    \setkeys{Gin}{height=0.9\textheight}
    \setkeys{Gin}{keepaspectratio=true}
    \graphicspath{{./figs/}}

\usepackage{float}
\usepackage{tikz}
\usepackage{tikzscale}
\usepackage{hyperref}
\makeatletter
\renewcommand*\aa@pageof{, page \thepage{} of \pageref*{LastPage}}
\makeatother

\usepackage{prettyref}  
\newrefformat{fig}{\text{Fig.~\ref{#1}}}
\newrefformat{eq}{\text{Eq.~\ref{#1}}}
\newrefformat{tab}{\text{Table~\ref{#1}}}
\newrefformat{sec}{\text{Sect.~\ref{#1}}}
\newrefformat{enu}{\text{~\ref{#1}}}

\newcommand{\pref}[1]{\prettyref{#1}}
\renewcommand{\href}{\ensuremath{h_{\text{ref}}}}
\begin{document}

\title{Self-scattering on large, porous grains in protoplanetary disks with dust settling}
\titlerunning{Self-scattering on porous grains}

\author{R. Brunngräber \and S. Wolf}
\institute{Institut für Theoretische Physik und Astrophysik, Christian-Albrechts-Universität zu Kiel, Leibnizstr. 15, 24118 Kiel, Germany\\\email{rbrunngraeber@astrophysik.uni-kiel.de}}

\date{Received / Accepted}

\abstract{Observations of protoplanetary disks in the sub-millimetre wavelength range suggest that polarisation is caused by scattering of thermal re-emission radiation. Most of the dust models that are used to explain these observations have major drawbacks: They either use much smaller grain sizes than expected from dust evolution models, or result in polarisation degrees that are lower than observed.}
{We investigate the effect of dust grain porosity on the observable polarisation due to scattering at sub-millimetre wavelengths arising from grain size distributions up to millimetre sizes, as they are expected to be present close to the midplane of protoplanetary disks.}
{Using the effective medium theory, we calculated the optical properties of porous dust and used them to predict the behaviour of the observable polarisation degree due to scattering. Subsequently, Monte Carlo radiative transfer simulations for protoplanetary disks with porous dust grains were performed to analyse the additional effect of the optical depth structure, and thus the effect of multiple scattering events and inhomogeneous temperature distributions on the net observable polarisation degree.}
{We find that porous dust grains with moderate filling factors of about 10\,\% increase the degree of polarisation compared to compact grains. For higher grain porosities, that is, grains with a filling factor of 1\,\% or lower, the extinction opacity decreases, as does the optical depth of a disk with constant mass. Consequently, the unpolarised direct radiation dominates the scattered flux, and the degree of polarisation drops rapidly. Even though the simulated polarisation degree is higher than in the case of compact grains, it is still below the typical observed values for face-on disks. However, the polarisation degree can be increased when crucial model assumptions derived from disk and dust evolution theories, for instance, dust settling and millimetre-sized dust grains, are dropped. In the case of inclined disks, however, our reference model is able to achieve polarisation degrees of about 1\,\%, and using higher disk masses, even higher than this.}
{}

\bibpunct{(}{)}{;}{a}{}{,}

\maketitle



\section{Introduction}
\label{sec:intro}
    Observations of the polarised continuum radiation of protoplanetary disks in the (sub-)millimetre (mm) wavelength range are considered an important source of information about the structure of magnetic fields and the dust grain shape and size in these disks. In this context, two main polarisation mechanisms are important: Dichroic emission and absorption by magnetically aligned non-spherical grains, and polarisation due to scattering. Especially the so-called self-scattering mechanism, that is, scattering of thermal re-emission radiation by large dust grains at (sub-)mm wavelengths, provides a promising explanation for many polarisation patterns that have recently been observed in protoplanetary disks \citep[e.g.][]{harris-et-al-2018,lee-et-al-2018,mori-et-al-2019,ohashi-kataoka-2019,sadavoy-et-al-2019}. The local polarisation degree in these spatially resolved maps has been found to be typically about \SIlist{1;3}{\percent,} with some rare outliers in both directions, for example, HD\,142527, VLA\,1623, or AS\,209 \citep{kataoka-et-al-2016,harris-et-al-2018,mori-et-al-2019}. To reproduce this relatively high value, it is often stated that dust grains with radii of about \SI{100}{\um} contribute most to the observed flux \citep{bacciotti-et-al-2018,hull-et-al-2018,lin-et-al-2020,ohashi-et-al-2020}. However, this explanation appears to contradict commonly accepted scenarios of grain growth in protoplanetary disks. There, grains are expected to grow rather fast to $\text{about}$ mm sizes and eventually dominate the emission in this wavelength regime \citep[e.g.][]{beckwith-sargent-1991,testi-et-al-2014}. Recent radiative transfer simulations allowing such large grains, however, predict lower polarisation degrees than observed \citep{brunngraeber-wolf-2020,yang-li-2020}. Furthermore, due to size-dependent turbulent vertical mixing, larger grains will settle towards the midplane, which leads to even lower polarisation degrees \citep{brunngraeber-wolf-2020}. This significant discrepancy in the degree of polarisation between simulation and observation is a major drawback in the explanation of the observed polarisation with scattering.

    To overcome this lack of polarisation in simulations, more complex dust models have been proposed, such as oblate, prolate, or fractal grain shapes, or different chemical compositions \citep{kirchschlager-bertrang-2020,brunngraeber-wolf-2020,yang-li-2020}. In the quoted studies, the polarisation after single scattering of an initially unpolarised wave is usually discussed, that is, the ratio of the Müller matrix elements $\nicefrac{S_{12}}{S_{11}}$ and the albedo $\omega$. However, these studies lack full radiative transfer simulations to prove the applicability of these results in the environment of protoplanetary disks, for instance. The effects of multi-scattering, attenuation of the radiation, and the contribution of unpolarised re-emission radiation are likely to change the degree of polarisation drastically.

    A particularly frequently proposed solution for the discrepancy of the degree of polarisation is the use of porous dust grains because the polarisation after single scattering increases with increasing porosity \citep[e.g.][]{tazaki-et-al-2019,brunngraeber-wolf-2020}. Theoretical and laboratory studies of dust growth and evolution both suggest low filling factors $\Phi$, that is, highly porous grains, and/or fractal geometries \citep{blum-et-al-2000,blum-wurm-2008,kothe-et-al-2013,garcia-gonzalez-2020}. Furthermore, dust particles collected in the Solar System and the upper layers of the Earth's atmosphere show a fractal and/or porous structure as well \citep{brownlee-et-al-1977,ahearn-et-al-2005,abe-et-al-2006,westphal-et-al-2014}. In this paper, we examine the effect of dust grain porosity on the polarised fraction of scattered radiation at sub-mm wavelengths of protoplanetary disks. Therefore we focus at first on the optical properties that determine the level of polarised scattered radiation in \pref{sec:opt_prop}. This approach is commonly used because it is independent of any underlying density distribution and thus requires no elaborate radiative transfer simulations. At the same time, it allows a straightforward examination of a broad parameter space. The dust model is given by a grain size distribution composed of porous dust, and the effect of different filling factors and maximum grain sizes is investigated. This dust model is described in \pref{sec:dust_model}, and the results are presented in \pref{sec:res_opt_prop} and discussed in \pref{sec:concl_opt_prop}. In the second part of this paper, in \pref{sec:rad_trans_sim}, radiative transfer simulations are performed in the context of a protoplanetary disk where the largest grains have settled towards the midplane. For this purpose, we introduce the underlying disk model in \pref{sec:set-up}, and present the results of the Monte Carlo simulations in \pref{sec:res_simu}. The discussion of these results is provided in \pref{sec:discussion}, and we summarize the entire paper in \pref{sec:summary}.

\section{Porosity and the general effect on polarised scattered radiation}
\label{sec:opt_prop}
    As a first step to investigate the effect of grain porosity on the degree of polarisation due to scattering, we examine selected optical quantities of the dust. This procedure is independent of any spatial density distribution, therefore it is universal and only depends on the intrinsic attributes of the dust grains determining its optical properties, such as chemical composition, shape, internal structure, and refractive index. This approach is therefore also applied in many studies concerning scattering at (sub-)mm wavelengths to explore the parameter space of dust composition and maximum grain size in the size distribution with regard to the expected degree of polarisation \citep[e.g.][]{kataoka-et-al-2015,tazaki-et-al-2019,yang-li-2020,kirchschlager-bertrang-2020,brunngraeber-wolf-2020}.

    \subsection{Dust model}
    \label{sec:dust_model}
        Owing to growth and destruction processes in protoplanetary disks, dust grains are expected to have sizes that range over many orders of magnitude. Each grain size has a different wavelength-dependent absorption, emission, and scattering behaviour. To account for this, we integrated these quantities weighted by the size-dependent abundance over the entire grain size distribution, which was adopted to be a power law with $n(s)\propto s^{q}$ as found for the interstellar medium \citep{mathis-et-al-1977}, with $s$ being the radius of a spherical grain. If not stated otherwise, the size exponent $q$ was set to $\num{-3.5}$. The lower boundary of the size distribution is $s_{\text{min}}=\SI{5}{\nm}$ and the upper boundary $s_{\text{max}}$ is a free parameter.

        The dust model was chosen to consist only of spherical grains. For the chemical composition, we used the refractive indices of astronomical silicate (astrosil) and the two main orientations of graphite (parallel and perpendicular, \citealt{draine-malhotra-1993}) from \citet{draine-2003c}. To mimic porous dust grains in our study, we considered spherical grains with different filling factors $\Phi$ and explored their effect on the scattering polarisation. The filling factor is defined as the ratio of the volume of the dust material $V_{\text{dust}}$ to the total grain volume $V_{\text{total}}$,
        \begin{equation}
        \label{eq:filling}
            \Phi = 1-\mathcal{P} = \frac{V_{\text{dust}}}{V_{\text{total}}}\ ,
        \end{equation}
        with the porosity $\mathcal{P}$, with $0 \leq \mathcal{P} \leq 1$, that is, a filling factor of $\Phi=1$ represents compact grains.

        The complex refractive indices of these porous grains were approximated using the effective medium theory (EMT) and the Bruggeman mixing rule \citep{bruggeman-1935}. The optical quantities, that is, cross-sections of absorption and scattering $C_{\text{abs}}$ and $C_{\text{sca}}$, and the elements of the Müller matrix $\mathcal{S}$, were calculated with the approximation solution of the Mie theory \citep{mie-1908} using the tool \texttt{miex} \citep{wolf-voshchinnikov-2004}. These properties were calculated for grain size distributions with \num{101} logarithmically spaced different maximum grain sizes $s_{\text{max}}$ between \SI{1}{\um} and \SI{10}{\cm}, and for \num{101} logarithmically spaced different filling factors between \num{e-4} and \num{1}. Each of these \num{10201} calculations used \num{100} grain size bins between $s_{\text{min}}=\SI{5}{\nm}$ and $s_{\text{max}}$ to average the optical quantities over the size distribution \citep[their Eq.\,28--30]{wolf-voshchinnikov-2004}, and \num{361} scattering angles in the range $\left[\SI{0}{\degree},\SI{180}{\degree}\right]$.
    \subsection{Optical quantities}
    \label{sec:res_opt_prop}
        In this section, we restrict the discussion to astrosil because it is the most abundant component in protoplanetary disks. However, our general qualitative conclusions with respect to the effect of porosity on the polarisation characteristics are also applicable for the case of graphite; see \pref{sec:appendix_opt_prop} for the corresponding plots for graphite.

        \textbf{Single scattering:} First, we assumed an unpolarised wave, which is scattered exactly once by a spherical dust grain. In this simple case, the polarisation degree of the outgoing wave is only a function of the two Müller matrix elements $S_{11}$ and $S_{12}$ of the dust,
        \begin{equation*}
            p = -\frac{S_{12}}{S_{11}}\ .
        \end{equation*}
        We refer to $p$ as the single scattering polarisation. It is the most commonly used quantity to describe and predict the scattering polarisation of different dust grains \citep[e.g.][]{kataoka-et-al-2015,tazaki-et-al-2019,yang-li-2020,kirchschlager-bertrang-2020,brunngraeber-wolf-2020}. For spherical grains, the thermal re-emission of the dust is unpolarised and the highest contribution to the net, that is, observable, polarised flux is indeed given by single-scattering events. Hence, the quantity $p$ is a good qualitative indicator for the expected net polarisation degree. The Müller matrix elements and thus $p$ depend on the scattering angle $\theta$. The predominant scattering angle is a function of the actual geometrical set-up and wavelength $\lambda$.

\begin{figure*}
            \includegraphics[width=0.49\linewidth, height=0.45\textheight]{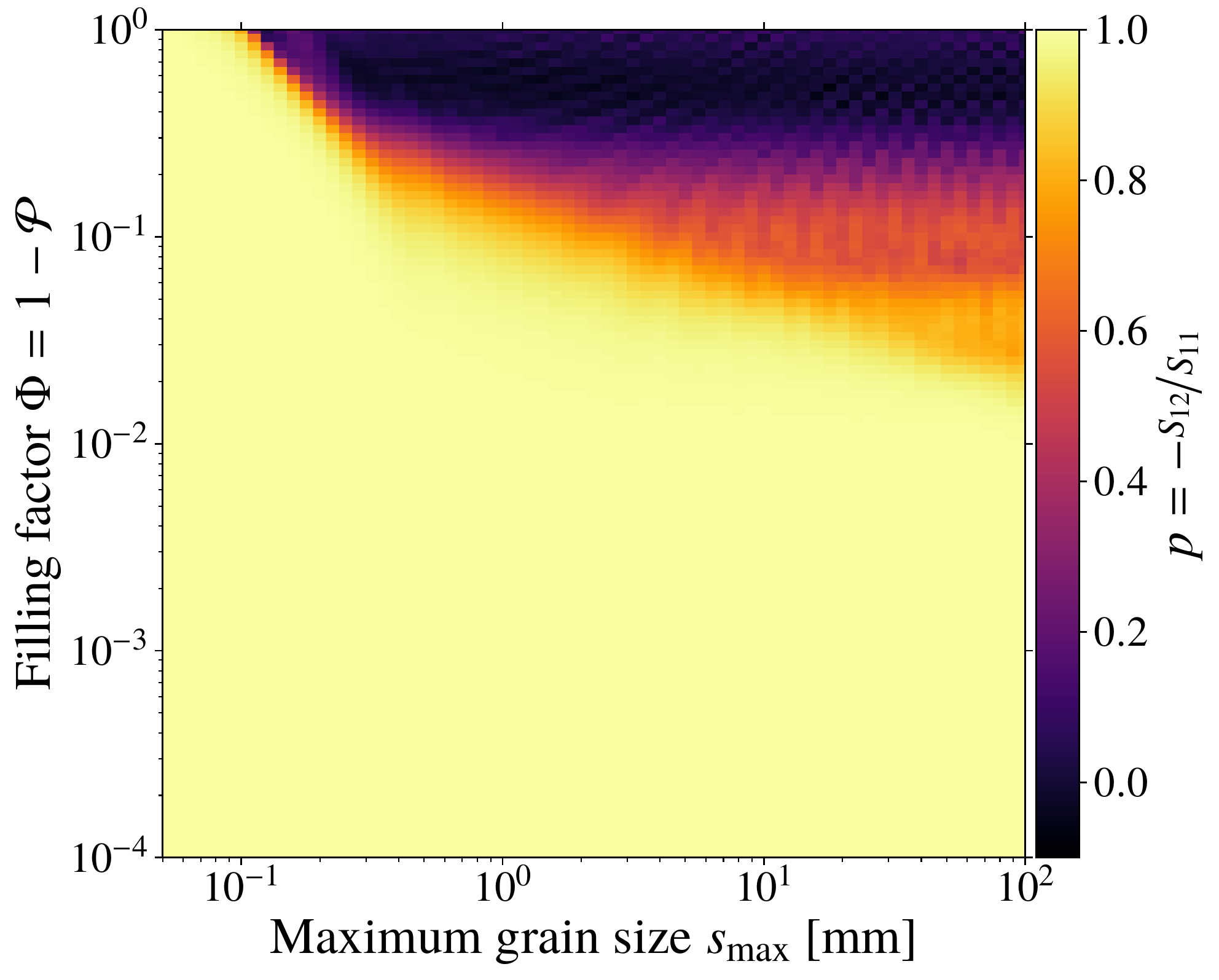}
            \includegraphics[width=0.49\linewidth, height=0.45\textheight]{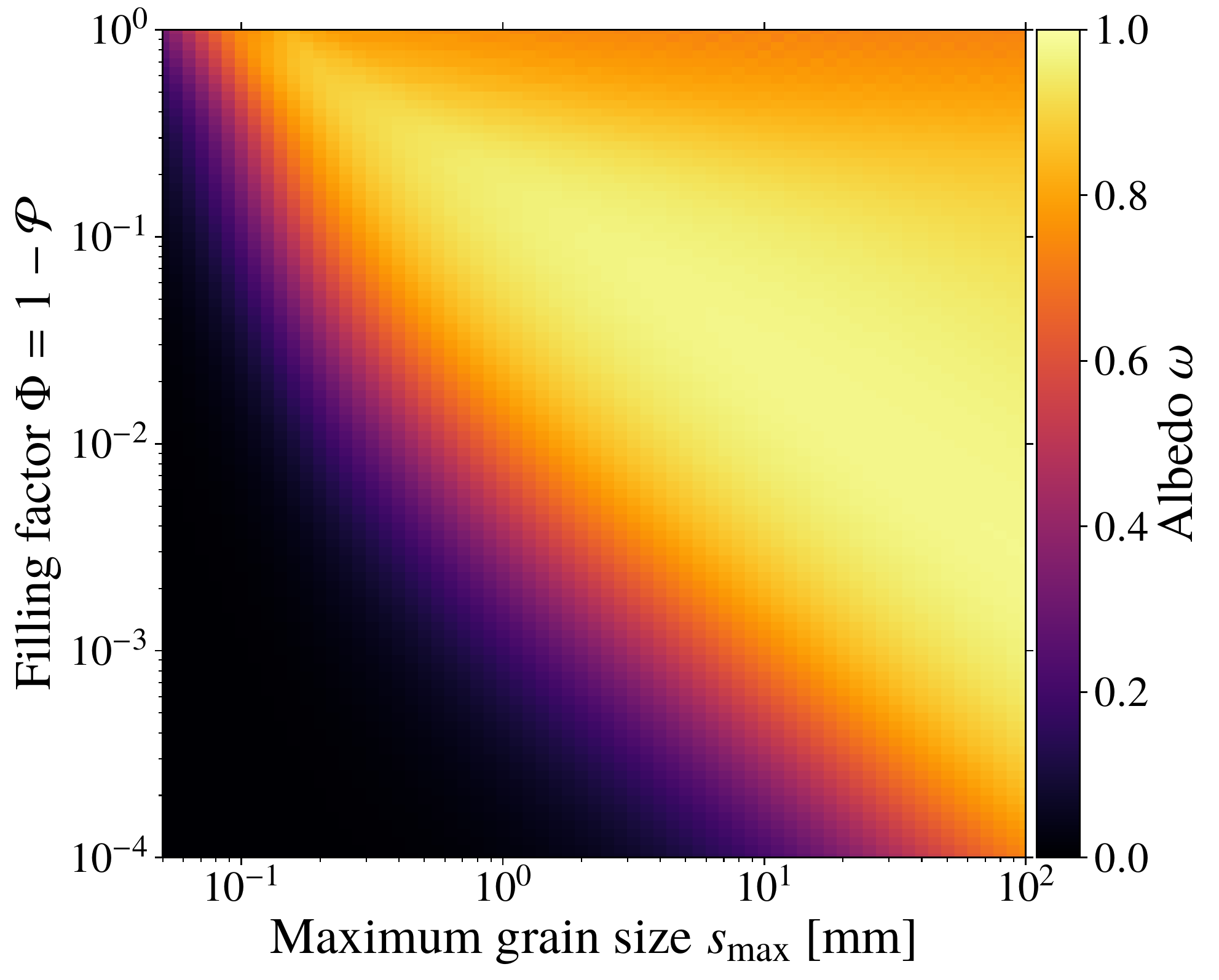}\\
            \includegraphics[width=0.49\linewidth, height=0.45\textheight]{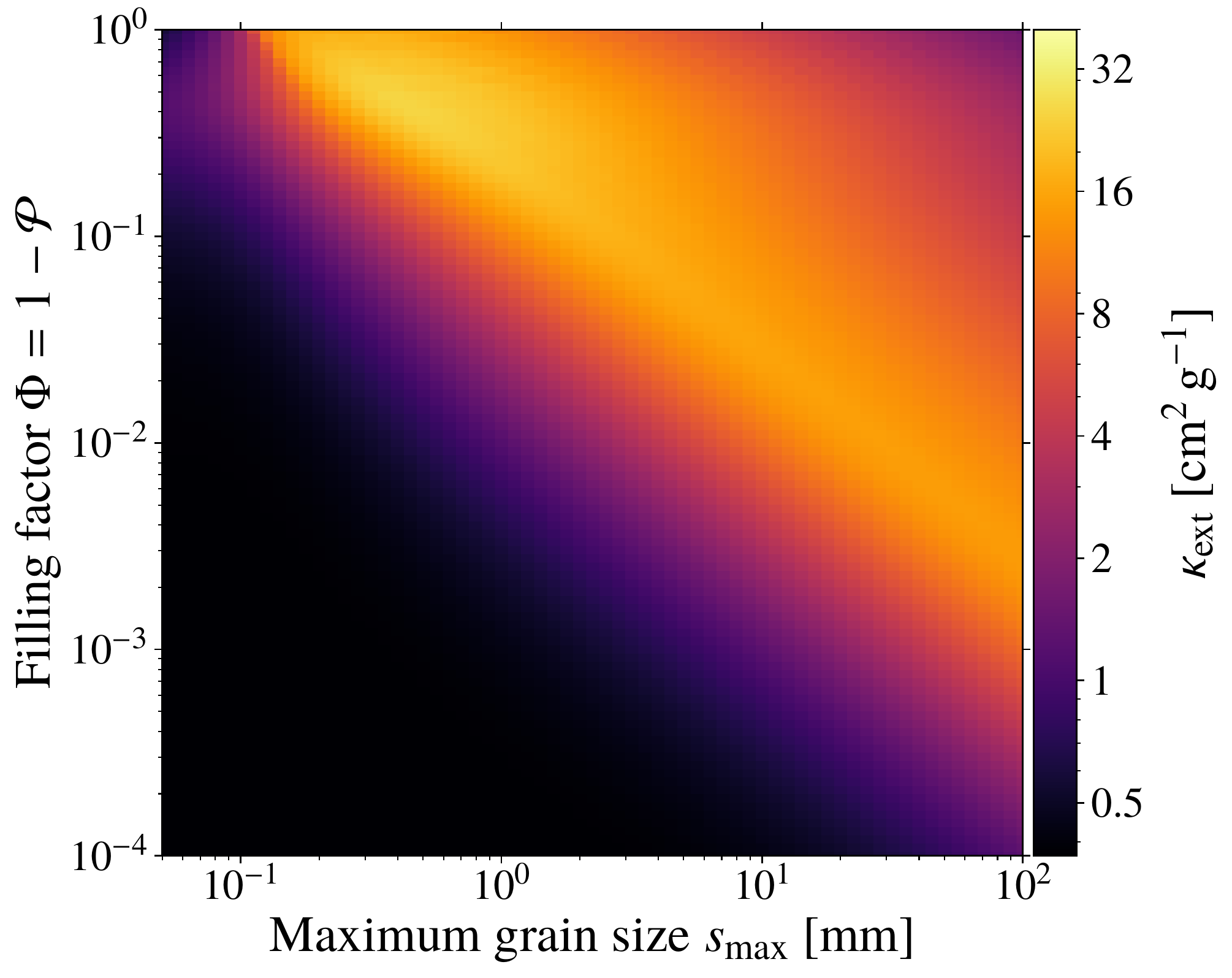}
            \includegraphics[width=0.49\linewidth, height=0.45\textheight]{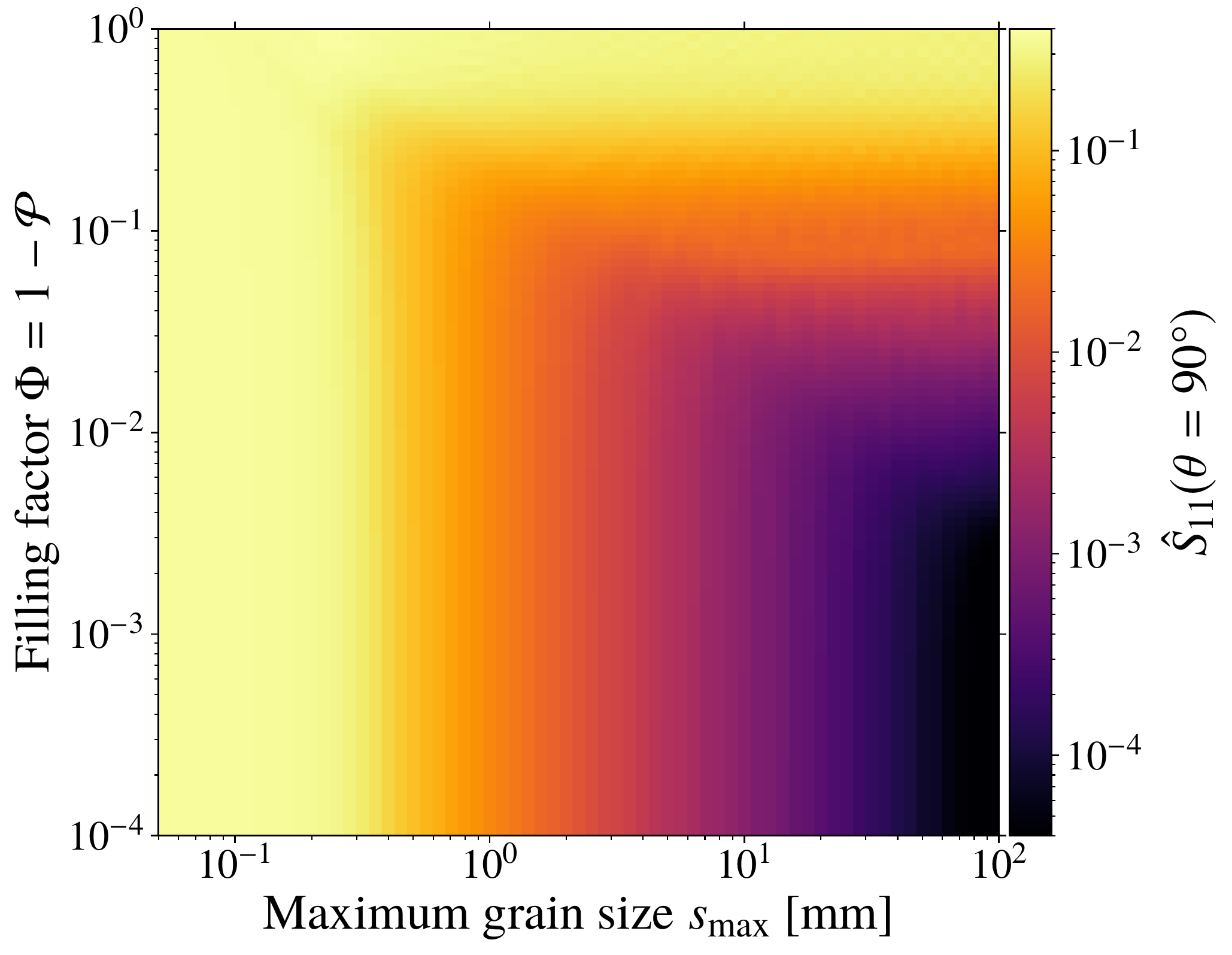}
            \caption{Degree of polarisation for single scattering $p = -\nicefrac{S_{12}}{S_{11}}$ for a scattering angle of $\theta=\SI{90}{\degree}$ \textit{(upper left)}, albedo $\omega$ \textit{(upper right)}, extinction opacity \textit{(lower left)}, and probability density function $\hat{S}_{11}$ for the scattering angle $\theta=\SI{90}{\degree}$ \textit{(lower right)}. All quantities are calculated for a grain size distribution with $q=\num{-3.5}$, $s_{\text{min}}=\SI{5}{\nm}$ and different upper grain size limits $s_{\text{max}}$ at a wavelength of \SI{850}{\um} for astrosil.}
            \label{fig:fill_smax_850}
        \end{figure*}

        For the Rayleigh scattering regime ($2\pi s \ll \lambda$), the single-scattering polarisation is maximised for a scattering angle of $\theta=\SI{90}{\degree}$. For larger grains (or smaller wavelengths), the single-scattering polarisation $p$ shows a fast-changing wavy pattern for different scattering angles \citep{brunngraeber-wolf-2019}. We therefore focused on a scattering angle of $\theta=\SI{90}{\degree}$. In addition, scattering angles close to this value also dominate the radiation of face-on protoplanetary disks, as shown in \pref{sec:rad_trans_sim}. The corresponding single-scattering polarisation $p$ is shown in the upper left panel of \pref{fig:fill_smax_850} for grain size distributions with different maximum grain sizes between \SI{1}{\um} and \SI{10}{\cm} and filling factors between \num{e-4} and \num{1} at a wavelength of $\lambda=\SI{850}{\um}$. We restrict the discussion here to this wavelength, which is close to the wavelength of many of the polarisation observations performed with ALMA \citep[e.g.,][]{stephens-et-al-2017,hull-et-al-2018,lee-et-al-2018,dent-et-al-2019,lin-et-al-2020}. Almost the entire parameter space yields a single-scattering polarisation of unity. The polarisation is thus determined by Rayleigh scattering although the grain sizes are partially much larger than the wavelength. Only for high filling factors (low porosities) is the degree of polarisation lower as a result of Mie scattering.
        \textbf{Scattering probability:} However, in contrast to the above simplistic scenario, the total flux does not only include the scattered radiation, but also the direct thermal emission of dust, that is, unscattered radiation. As this direct emission is unpolarised according to our dust model, the net polarisation degree is lower than the polarisation degree resulting from scattering alone. In order to derive a useful measure for the total net polarisation degree under these more complex circumstances, we have to quantify whether a photon interacts with dust grains (scattering or absorption) on its path to the observer. Depending on the optical depth $\tau \propto \kappa_{\text{ext}} \cdot \rho_{\text{dust}}$ along that path, the probability of interaction is given by $1-\exp{(-\tau)}$ with the extinction opacity $\kappa_{\text{ext}} = \nicefrac{C_{\text{ext}}}{m_{\text{dust}}}$, the extinction cross-section $C_{\text{ext}} = C_{\text{sca}} + C_{\text{abs}}$, the mean grain mass $m_{\text{dust}}$, and the spatial dust mass density $\rho_{\text{dust}}$. If the observed density distribution is optically thin ($\tau\ll1$), the observable flux is determined by unpolarised direct thermal re-emission radiation. For a given optical depth, the ratio of scattered radiation to direct thermal radiation is also determined by the fraction of scattering events among all interactions. Consequently, the polarisation degree of the emerging radiation is given by both the optical depth and the albedo $\omega$,
        \begin{equation*}
            \omega = \frac{C_{\text{sca}}}{C_{\text{sca}} + C_{\text{abs}}}\ .
        \end{equation*}
        The albedo and extinction opacity, which determines the optical depth, are shown in the upper right and lower left panel of \pref{fig:fill_smax_850}, respectively. For large maximum grain sizes, the opacity first increases with decreasing filling factor and decreases again after reaching a maximum value. Whereas the grain mass is proportional to the filling factor, the extinction cross-section shows a more complex behaviour. At first, it decreases much slower with decreasing filling factor than the grain mass. However, for filling factors $\Phi\lesssim\num{e-2}$, the cross-section is roughly proportional to $\Phi^2$ and thus decreases faster than the grain mass. Hence, the opacity is maximised at the transition of these two regions. In addition, the opacity is constant for a constant product $s_{\text{max}}\cdot\Phi$, which has been shown previously by \citet{kataoka-et-al-2014} and \citet{tazaki-et-al-2019}, for example.
        \textbf{Scattering angle dependency:} If a scattering event occurs, the actual intensity of the radiation scattered in the direction of the observer is also a function of the scattering angle $\theta$. For single scattering, this dependency is given by the first Müller matrix element $S_{11}(\theta)$. The probability density function $\hat{S}_{11}(\theta) = \frac{S_{11}\left(\theta\right)}{\int_{0}^{\pi}{S_{11}\sin{\theta}\,\text{d}\theta}}$ for a scattering angle of $\theta=\SI{90}{\degree}$ is shown in the lower right panel of \pref{fig:fill_smax_850}. For large maximum grain sizes, this probability decreases for increasing grain size because large grains tend to scatter predominantly in the forward direction, that is, $\theta\approx\SI{0}{\degree}$. With increasing grain size, the intensity of the radiation that is scattered at $\SI{90}{\degree}$ decreases. As a consequence, the flux ratio of scattered to direct radiation and hence the degree of polarisation decreases as well.

    \subsection{Conclusion}
    \label{sec:concl_opt_prop}
\begin{figure}
            \includegraphics[width=0.9\linewidth, height=0.45\textheight]{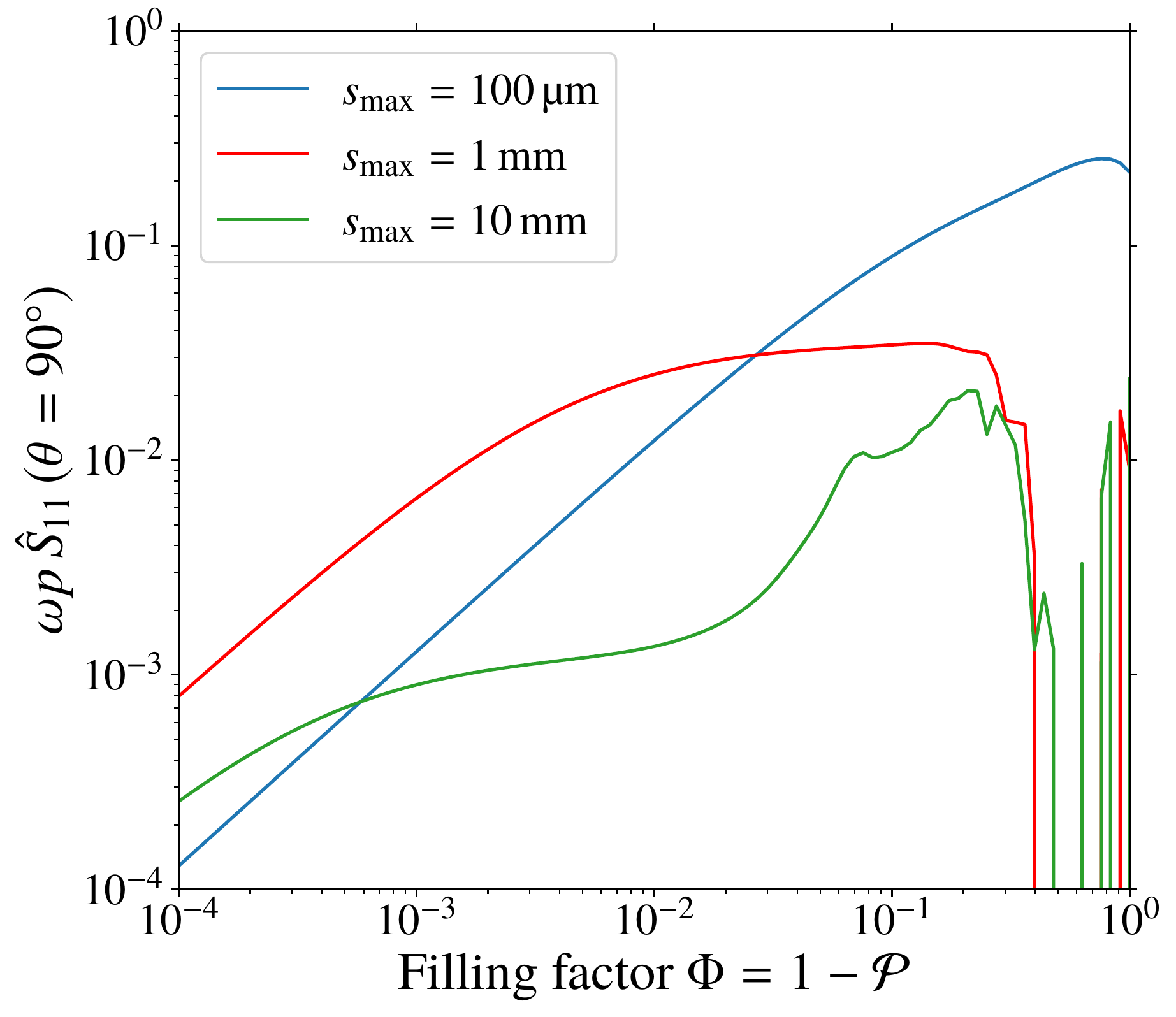}
            \caption{Product of albedo $\omega$, single-scattering polarisation $p$, and probability density function $\hat{S}_{11}$ for the scattering angle $\theta=\SI{90}{\degree}$. All quantities are calculated for a grain size distribution with $q=\num{-3.5}$, $s_{\text{min}}=\SI{5}{\nm}$ and three different maximum grain sizes $s_{\text{max}} = \SI{100}{\um}$ (\textit{blue}), $\SI{1}{\mm}$ (\textit{red}), and $\SI{10}{\mm}$ (\textit{green}) at a wavelength of \SI{850}{\um} for astrosil; see \pref{sec:concl_opt_prop} for the corresponding discussion.}
            \label{fig:pol_product_fill}
        \end{figure}

        To evaluate the effect of the grain porosity of the observable polarisation, a combined analysis of all of the aforementioned quantities is mandatory. Lower filling factors increase the single-scattering polarisation $p$ and the albedo $\omega$. At the same time, however, the extinction opacity, and thus the scattering probability, decreases as well as the probability $\hat{S}_{11}$ for the scattering angle $\theta=\SI{90}{\degree}$. In \pref{fig:pol_product_fill}, the product of albedo, single-scattering polarisation, and the scattering probability density function for $\theta=\SI{90}{\degree}$ as a function of the filling factor $\Phi$ is shown for different maximum grain sizes $s_{\text{max}}$. This product can be used as an advanced proxy for the polarisation degree due to scattering for optically thick environments. For grains in the Rayleigh scattering regime, a decrease in filling factor $\Phi$ decreases this proxy and therefore the expected polarisation degree. For size distributions with large maximum grain sizes compared to the wavelength and high filling factors, the usual wavy pattern of the Mie scattering regime is present, which also decreases this quantity. Thus, for the combination of sub-mm wavelengths and large grain sizes, the degree of polarisation is expected to first increase with decreasing filling factor and after reaching a maximum value, to decrease again for even lower filling factors.

        In summary, the analysis of these optical quantities derived from the complex refractive index already provides useful results in the context of polarisation due to (sub-)mm scattering off porous dust grains. First, increasing the porosity, that is, decreasing the filling factor, results in a higher single-scattering polarisation $p$ of up to \SI{100}{\percent}. In addition, the albedo increases up to unity as well. Thus, the quantity $p\cdot\omega$, which is occasionally used in the literature as a simple proxy for expected polarisation degrees \citep[e.g.][]{kataoka-et-al-2015,tazaki-et-al-2019,yang-li-2020}, is maximised.

        Second, highly porous grains tend to scatter almost exclusively in the forward direction. For the considered spherical grains, forward scattering does not produce polarised radiation. This reduces the scattering probability and polarised flux for larger scattering angles.

        Third, above a certain threshold of the maximum grain size of about \SI{0.3}{\mm} for the size distribution, composition, and wavelength considered in \pref{fig:fill_smax_850}, the extinction opacity first increases with increasing filling factors but decreases after reaching a maximum value. For filling factors below this maximum, the opacity drops rather fast and the optical depth decreases. This reduces the ratio of scattered to direct radiation and ultimately, the degree of polarisation. In addition, the opacity is roughly constant for a constant product $s_{\text{max}}\cdot\Phi$.

        And lastly, considering the single-scattering polarisation $p$ alone is in general not sufficient to predict the net polarisation nor its qualitative behaviour, that is, trends as a function of grain properties. Three more quantities result from the refractive index of the dust and contribute significantly to the polarisation degree: albedo, extinction opacity, and the fraction of radiation scattered in the direction of the observer.
\section{Radiative transfer simulations of a protoplanetary disk}
\label{sec:rad_trans_sim}
    In order to evaluate the effect of dust grain porosity on the scattering polarisation in the sub-mm wavelength regime in the environment of protoplanetary disks, comprehensive numerical simulations are mandatory to account for the complex processes that determine the observable polarisation. Therefore full 3D Monte Carlo radiative transfer simulations with the publicly available\footnote{\url{http://www1.astrophysik.uni-kiel.de/~polaris/index.html}} software \texttt{POLARIS} \citep{reissl-et-al-2016} were conducted. In addition to the optical properties that were used in the simple approach in \pref{sec:opt_prop}, these simulations also considered self-consistent dust temperature for each grain size bin, the contribution of unpolarised direct re-emission, multiple scattering, local radiation field anisotropy, the scattering angle as a function of spatial density distribution and viewing angle, and a size-dependent vertical density distribution resulting from dust settling. For this purpose, we first define the density distribution of our protoplanetary disk model including dust settling in \pref{sec:set-up}. In \pref{sec:res_simu}, the results of the radiative transfer simulations are discussed.

    \subsection{Model set-up}
    \label{sec:set-up}
        For high polarisation due to scattering, a high optical depth is necessary. Therefore self-scattering in the \si{mm} wavelength regime is almost exclusively observed in protoplanetary disks. In these disks, grain sizes from a few nanometres up to centimetres are expected to exist \citep[e.g.][]{beckwith-sargent-1991,testi-et-al-2014}. Due to gas drag, the larger grains settle towards the midplane of the disk, whereas the smallest grains are coupled to the gas and thus show a much larger vertical extent \citep{safronov-1969,goldreich-ward-1973,dubrulle-et-al-1995,dullemond-dominik-2004b,woitke-et-al-2016}. The radial and vertical density distribution of the dust is therefore a function of the grain size \citep{graefe-et-al-2013,pinte-et-al-2016,guilloteau-et-al-2016,avenhaus-et-al-2018,andrews-et-al-2018,villenave-et-al-2020}. Additionally, the scattering and polarisation efficiency also depend on grain size. To account for these effects, the radiative transfer simulations in this study were performed on the basis of a protoplanetary disk model where the largest grains have already settled towards the midplane. A more detailed description of the disk model can be found in \citet{brunngraeber-wolf-2020}. The gas density distribution $\rho_{\text{gas}}$ is given by \citep{lynden-bell-pringle-1974,kenyon-hartmann-1995,hartmann-et-al-1998}
        \begin{equation*}
            \rho_{\text{gas}}(r,z) = \frac{\Sigma_{\text{gas}}(r)}{\sqrt{2\pi}\,h_{\text{gas}}(r)}\!\cdot\exp{\left[-\frac{1}{2}\left(\frac{z}{h_{\text{gas}}(r)}\right)^2\right]}\ ,
        \end{equation*}
        with the vertically integrated surface density
        \begin{equation}
        \label{eq:sigma}
            \Sigma_{\text{gas}}(r) = \sqrt{2\pi}\,\rho_0\,\href{}\cdot\left(\frac{r}{R_0}\right)^{-\gamma}\cdot\exp{\left[-\left(\frac{r}{R_{\text{trunc}}}\right)^{2-\gamma}\right]}
        \end{equation}
        and the scale height
        \begin{equation*}
            h_{\text{gas}}(r) = \href{}\,\left(\frac{r}{R_0}\right)^{\beta}\ ,
        \end{equation*}
        with $(r,z)$ being the usual cylindrical coordinates, $R_0$ is the radius where the gas scale height $h_{\text{gas}}$ reaches the reference scale height \href{}, $R_{\text{trunc}}$ is the truncation radius, and $\rho_0$ is the density scaling parameter that defines the total gas mass.

        To account for the dust settling, the dust mass was re-distributed vertically. The dust scale height $h_{\text{dust}}(r,s,\Phi)$ was downscaled according to the settling model of \citet{dubrulle-et-al-1995} and \citet{woitke-et-al-2016} for grains with radius $s$ and bulk density $\rho_{\text{bulk}}$,
        \begin{equation}
        \label{eq:scale_height}
            h_{\text{dust}}(r,s,\Phi) = h_{\text{gas}}(r)\cdot \sqrt{\frac{f(r,s,\Phi)}{1+f(r,s,\Phi)}}\ ,
        \end{equation}
        where the settling function $f(r,s,\Phi)$ is given by
        \begin{equation}
        \label{eq:settling}
            f(r,s,\Phi) = \frac{\alpha_{\text{settle}}}{\sqrt{6\pi}}\ \frac{\Sigma_{\text{gas}}(r)}{s\,\varrho_{\text{bulk}}\left(\Phi\right)}\ .
        \end{equation}
        The parameter $\alpha_{\text{settle}}$ regulates the strength of the settling and is determined by the disk viscosity and the strength of turbulent mixing \citep{dubrulle-et-al-1995,woitke-et-al-2016}. We wish to emphasise that due to the dependency on $\varrho_{\text{bulk}}\left(\Phi\right),$ the strength of the vertical settling is thus a function of porosity and is weaker for highly porous grains. The mass density distribution of the dust is hence given by
        \begin{equation}
        \label{eq:dens_distro}
            \rho_{\text{dust}}(r,z,s,\Phi) = \frac{\Sigma_{\text{dust}}(r)}{\sqrt{2\pi}\,h_{\text{dust}}(r,s,\Phi)}\!\cdot\exp{\left[-\frac{1}{2}\left(\frac{z}{h_{\text{dust}}(r,s,\Phi)}\right)^2\right]}
        \end{equation}
        and the dust surface density by
        \begin{equation*}
            \Sigma_{\text{dust}}(r) = f_{\text{d/g}}\cdot\Sigma_{\text{gas}}(r)
        \end{equation*}
        with the dust-to-gas ratio $f_{\text{d/g}}=\nicefrac{1}{100}$.

        The dust model differs slightly from the model described in \pref{sec:dust_model}. Whereas in the previous section the optical quantities, for instance, the cross-sections, were averaged over the entire size distribution for simplicity, during the radiative transfer simulation, the size as well as the corresponding optical quantities of a grain were determined probabilistically for each individual interaction event (scattering or absorption). To achieve ample sampling, we used a total of \num{1000} logarithmically spaced grain sizes to represent the entire distribution. Furthermore, a mixture of \SI{62.5}{\percent} astrosil and \SI{37.5}{\percent} graphite ($\nicefrac{1}{3}$--$\nicefrac{2}{3}$ approximation; \citealt{draine-malhotra-1993}) was used.

        To account for dust settling, the size-dependent scale height given by \pref{eq:scale_height} and \pref{eq:settling} was calculated for ten logarithmically spaced size bins. These bins cover the entire range of grain sizes from $s_{\text{min}}$ to $s_{\text{max}}$. Thus, the net density distribution is given by superposition, that is, the sum of ten individual density distributions characterised by their individual grain size ranges; for details, see also \citet{brunngraeber-wolf-2020}. The parameter values of the reference disk model and the central star are identical to those of the reference model of \citep{brunngraeber-wolf-2020}; see \pref{tab:model_set-up}.

        \begin{table}
            \caption{Disk and stellar parameter values of our reference case.}
            \label{tab:model_set-up}
            \centering
            \begin{tabular}{l l l}
                \hline\hline
                \rule{0pt}{2.5ex}Parameter          & Variable                                  & Values    \\[1mm]
                \hline
                \rule{0pt}{2.5ex}Inner radius       & $R_{\text{in}}$ [\si{\au}]                & \num{0.1} \\
                Outer radius                        & $R_{\text{out}}$ [\si{\au}]               & \num{300} \\
                Reference radius                    & $R_{0}$ [\si{\au}]                        & \num{100} \\
                Truncation radius                   & $R_{\text{trunc}}$ [\si{\au}]             & \num{100} \\
                Reference scale height              & \href{} [\si{\au}]                        & \num{10} \\
                Density profile                     & $\gamma$                                  & \num{1.1} \\
                Flaring parameter                   & $\beta$                                   & \num{1.1} \\
                Dust mass                           & $M_{\text{dust}}$ [\si{\msun}]            & \num{e-4} \\
                Settling parameter                  & $\alpha_{\text{settle}}$                  & \num{e-2} \\
                Inclination                         & $i$ [\si{\degree}]                        & \num{0} \\
                Stellar luminosity                  & $L_{\star}$ [\si{\lsun}]                  & \num{0.9} \\
                Stellar temperature                 & $T_{\star}$ [\si{\K}]                     & \num{4050} \\[.8ex]
                \multirow{3}{30mm}{Dust composition (mass fraction)} &                          & Silicate: \SI{62.5}{\percent} \\
                                                    &                                           & Graphite$_{\perp}$: \SI{25}{\percent} \\
                                                    &                                           & Graphite$_{\parallel}$: \SI{12.5}{\percent} \\[.8ex]
                \multirow{3}{*}{Grain bulk density} & \multirow{3}{15mm}{$\varrho_{\text{bulk}}$ [\si{\kg\per\m\cubed}]}    & Silicate: \num{3800} \\
                                                    &                                           & Graphite: \num{2200} \\
                                                    &                                           & Mixture: $\sim \num{2986}$ \\[.8ex]
                Minimum grain size                  & $s_{\text{min}}$ [\si{\nm}]               & \num{5} \\
                Maximum grain size                  & $s_{\text{max}}$ [\si{\mm}]               & \num{1} \\
                Size distribution                   & $q$                                       & \num{-3.5} \\
                \hline
            \end{tabular}
        \end{table}
    \subsection{Results}
    \label{sec:res_simu}
        In this section, we present the outcome of the radiative transfer simulations. In order to derive spatially resolved polarisation maps, we first calculated the temperature distribution of the dust. The sole heating source is a central star with an effective temperature of $T_\star=\SI{4050}{\K}$ and a luminosity of $L_\star=\SI{0.9}{\lsun}$, representing a low-luminous but typical T Tauri star \citep[see e.g. the compilation in Table\,F.1. in][]{varga-et-al-2018}. Based on the derived self-consistent temperature distribution, the radiative transfer of the thermal re-emission of the dust and thus the resulting intensity maps were simulated at $\lambda=\SI{850}{\um}$, the same wavelength as in \pref{sec:opt_prop}. The observable flux represents a combination of the unpolarised direct emission and the scattered radiation. In the following, we separate these two individual contributions. To compare the results of models with different filling factors, we present radial profiles of the degree of polarisation both for the scattered and the total flux and the flux ratio of direct to scattered radiation. More details on the simulation process and extraction of the radial profiles can be found in \citet{brunngraeber-wolf-2020}.

\begin{figure*}
            \includegraphics{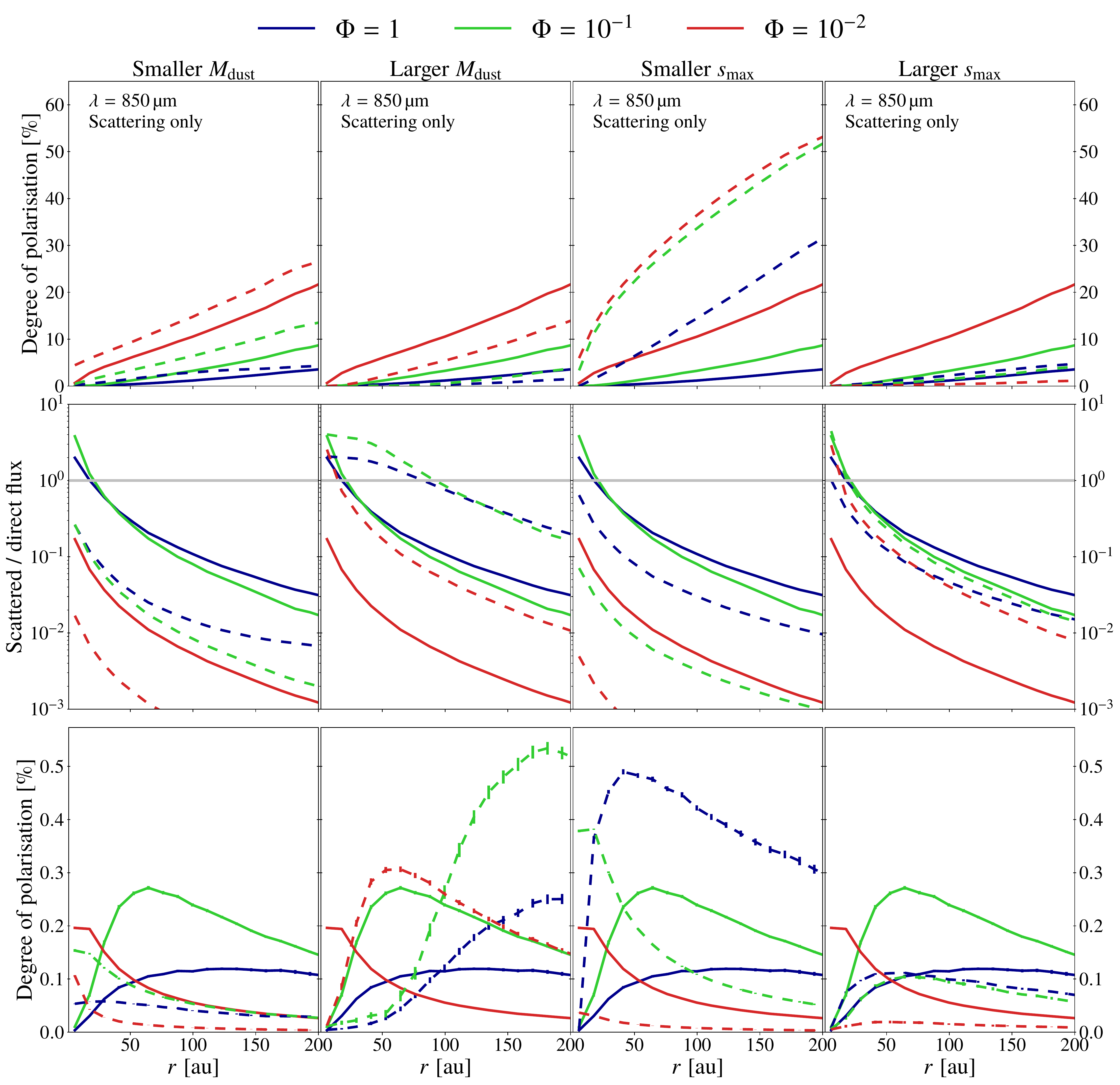}
            \caption{Radial profiles of different quantities for disks with different filling factors (colours). The reference disk model (solid lines) is shown in a comparison with different disk models (dashed lines): A disk with a dust mass $M_{\text{dust}} = \SI{e-5}{\msun}$ lower by a factor ten (\textit{first column}), a disk with a dust mass $M_{\text{dust}} = \SI{e-3}{\msun}$ higher by a factor ten (\textit{second column}), a disk with a maximum grain size $s_{\text{max}} = \SI{100}{\um}$ smaller by a factor ten (\textit{third column}), and a disk with a maximum grain size $s_{\text{max}} = \SI{10}{\mm}$ larger by a factor ten (\textit{fourth column}). \textit{Top}: Degree of polarisation for scattering only, i.e. without the unpolarised direct re-emission. \textit{Middle}: Flux ratio of scattered to direct re-emission. \textit{Bottom}: Net degree of polarisation of the total flux, i.e. scattered and direct radiation, including vertical error bars.}
            \label{fig:res_simu_mass_mass_amax_amax}
        \end{figure*}

        First, we focus on the scattered radiation. In this way, the comparison to the results of the simple approach in \pref{sec:opt_prop} is straightforward. In the upper row of \pref{fig:res_simu_mass_mass_amax_amax}, radial profiles of the polarisation degree for the scattered flux only for different filling factors $\Phi$ are shown. The polarisation of pure scattering increases with decreasing filling factor (different colours), which is a direct consequence of an increasing ratio of the Müller matrix elements $\nicefrac{S_{12}}{S_{11}}$; see also the upper left panel of \pref{fig:fill_smax_850}. In addition, the general shape of the radial profiles is determined by the degree of anisotropy of the local radiation field. An anisotropic radiation field is one of the main requirements for polarisation due to scattering of thermal re-emission radiation \citep{kataoka-et-al-2015,heese-wolf-brauer-2020,brunngraeber-wolf-2020}. In the inner regions of the disk, the radiation field is close to isotropic because of the axisymmetric density and temperature distribution. However, the local radiation field becomes increasingly anisotropic with larger orbital radii $r$. This is caused by the decreasing temperature, and hence flux, for larger distances. Consequently, the degree of polarisation increases towards the outer regions of the disk.

        Both a lower disk mass (first column in \pref{fig:res_simu_mass_mass_amax_amax}) and a smaller maximum grain size (third column) increase the polarisation of scattered flux. In both cases, the temperature in the upper layers of the disk is higher than in the reference case (solid lines in all four columns). The resulting steeper vertical temperature gradient causes the anisotropy of the local radiation field and thus the polarisation degree to increase. Additionally, for a smaller maximum grain size, the single-scattering polarisation $p$ is significantly higher than in the reference case; see the upper left panel of \pref{fig:fill_smax_850}. Similarly, the scattering polarisation decreases for increasing disk masses (second column) and maximum grain sizes (fourth column).

        Nonetheless, a high degree of scattering polarisation is not the only prerequisite for a high net polarisation degree combining scattered and direct radiation. The flux ratio of scattered to direct radiation is shown in the middle row of \pref{fig:res_simu_mass_mass_amax_amax}. A high flux ratio ($\gg1$) represents the case in which the total flux is dominated by scattering and thus the degree of polarisation is similar to the case with scattered flux only. However, for our simulations, the ratio is well below unity in most considered cases (down to $\sim\num{2e-2}$ for compact dust and even lower for porous grains), and thus the direct unpolarised radiation contributes much more to the overall flux than the scattered flux. In most of the cases shown in \pref{fig:res_simu_mass_mass_amax_amax}, the flux ratio decreases for a decreasing filling factor $\Phi$ because of two reasons. First, the extinction cross-section decreases for more porous grains. Second, the product of albedo, single-scattering polarisation, and probability for a scattering angle of $\SI{90}{\degree}$ also decreases for lower filling factors; see the lower left panels of \pref{fig:fill_smax_850} and \pref{fig:pol_product_fill}. In addition, for constant optical properties, the ratio of scattered to direct flux is essentially governed by the optical depth towards the observer. Thus, in the case of a lower or higher disk mass, the ratio decreases or increases, respectively, by the same factor as the mass. However, considering different maximum grain sizes, the flux ratio is also a function of the extinction opacity $\kappa_{\text{ext}}$ and the product of albedo, single-scattering polarisation, and probability distribution of the scattering angle. As a result, the flux ratio decreases by different amounts for most of the simulations with smaller and larger maximum grain sizes. One exception is the combination of a low filling factor of $\Phi=\num{e-2}$ and a maximum grain size of $s_{\text{max}}=\SI{10}{\mm}$ represented by the dashed red line in the fourth column of \pref{fig:res_simu_mass_mass_amax_amax}. Here, both the albedo and the extinction cross-section increase for the larger maximum grain size.

        The combination of the scattering polarisation degree and the flux ratio of scattered to direct radiation yields the overall degree of polarisation of the total flux; see the lower panels of \pref{fig:res_simu_mass_mass_amax_amax}. For most disk models, the net polarisation first increases with decreasing filling factor, and after reaching a maximum value, it decreases again. This is the result of two counteracting effects: On the one hand, there is the already mentioned steadily increasing degree of polarisation of pure scattering. On the other hand, there is the steep drop of the flux ratio with decreasing filling factors. The only exception within the considered parameter space is the disk with a smaller maximum grain size of $s_{\text{max}} = \SI{100}{\um}$. Here, the maximum polarisation is produced by compact grains and is reduced for more porous grains, which is also shown in \pref{fig:pol_product_fill}. Compared to the reference case, the net polarisation increases for a higher disk mass because the disk becomes optically thick and for a smaller maximum grain size due to the increase in scattering polarisation. In general, the net polarisation degree of the total flux is reduced significantly compared to the case of scattered radiation only. Furthermore, the shape of the radial profile changes and now decreases towards larger radii after a maximum degree of polarisation. Its position in the disk is governed by the ratio of the scattered to direct unpolarised flux. For optically thin disks, the maximum is located at the inner rim of the disk and is shifted farther out with increasing optical depth\footnote{assuming a constant albedo}. The position of maximum polarisation degree can therefore be used to determine the transition from optically thick to optically thin \citep{brunngraeber-wolf-2020}.

\begin{figure}
            \includegraphics[width=0.49\linewidth]{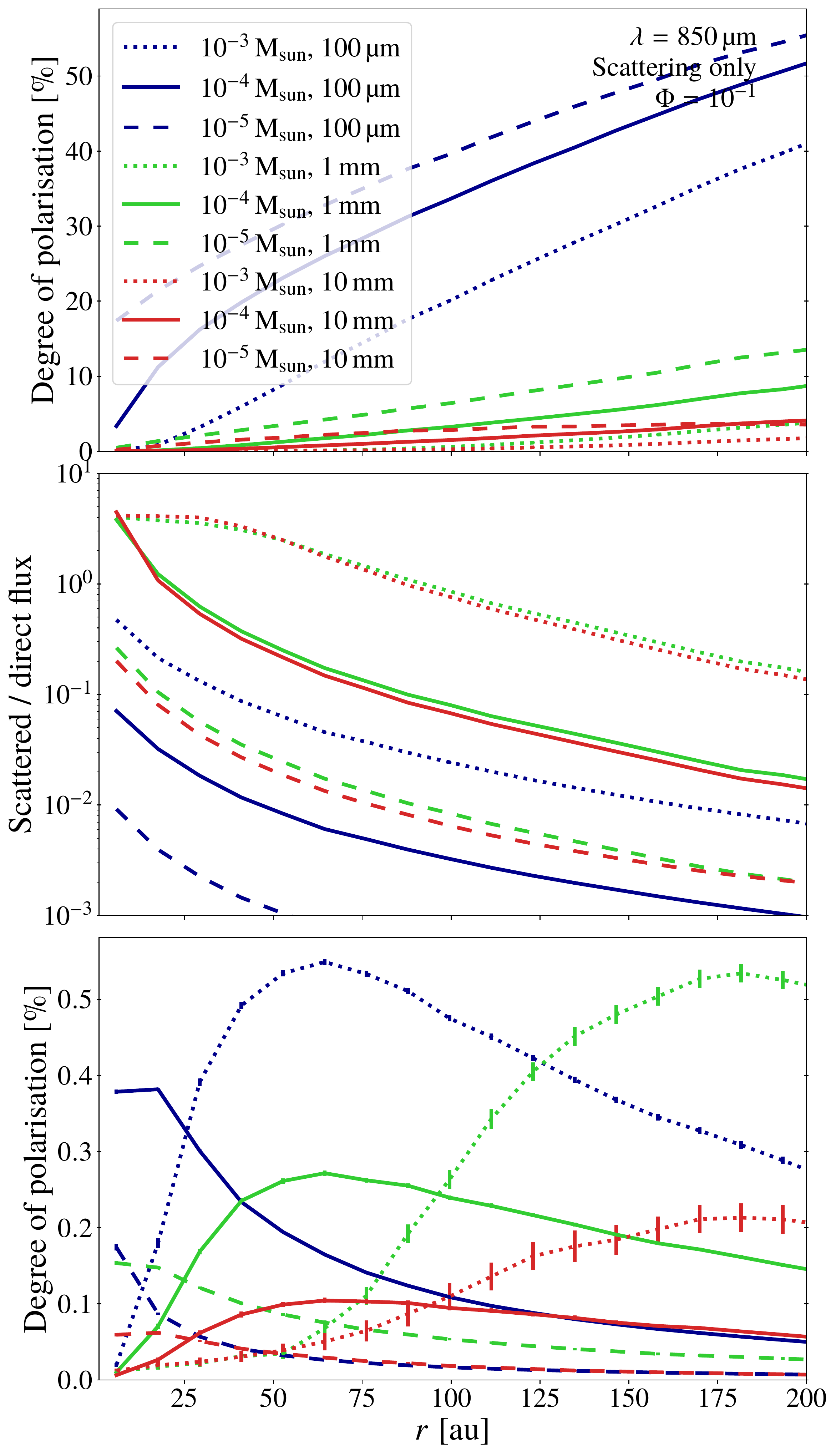}
            \includegraphics[width=0.49\linewidth]{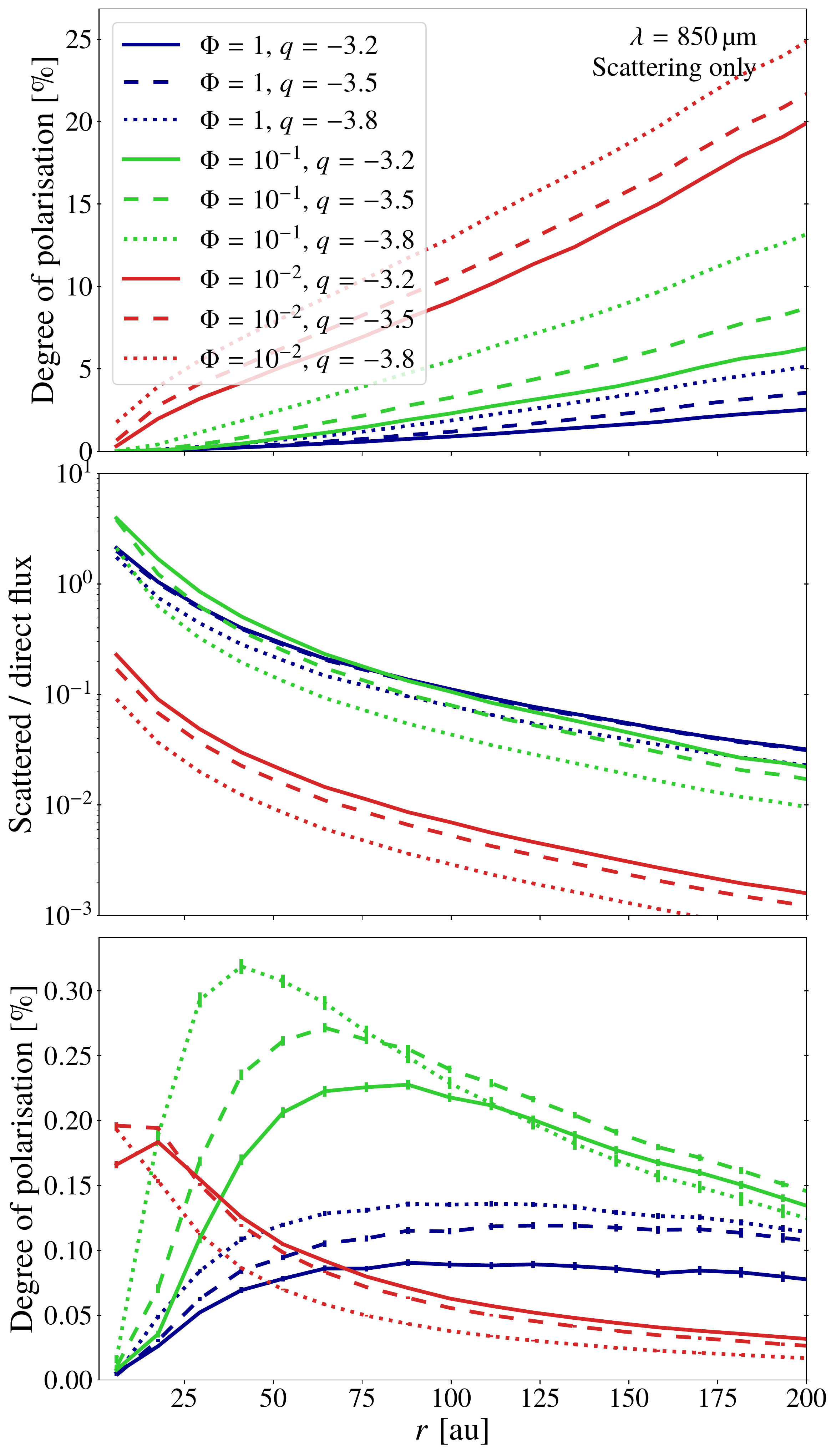}
            \caption{Same as \pref{fig:res_simu_mass_mass_amax_amax}, but for (\textit{left}) disks with different dust masses (line style) and upper grain size limits (colours), and for (\textit{right}) disks with different filling factors (colour) and exponents of the grain size distribution (line style).}
            \label{fig:res_simu_mass_amax_por_q}
        \end{figure}

        Although the degree of polarisation increases for porous grains, it is still significantly lower than those observed, which typically yield polarisation degrees of about a few percent. The main cause, as discussed, is the relatively low contribution of the scattered radiation to the total radiation. At first glance, an increase in this contribution and thus of the resulting polarisation degree can be achieved by assuming an increased maximum grain size, and thus a greater effect of large grains with large extinction cross-sections. However, our simulations show that this conclusion is not correct. In the left column of \pref{fig:res_simu_mass_amax_por_q}, the radial profiles for disks containing grains with different maximum radii and a filling factor of $\Phi=\num{e-1}$ are directly compared for three different total dust masses. For the chosen wavelength of $\lambda=\SI{850}{\um}$, the scattered to direct flux ratio is very similar for disks containing dust distributions with maximum radii of $s_{\text{max}} = \SI{1}{\mm}$ (green) and $\SI{10}{\mm}$ (red). This is because opacity and albedo are very similar as well. The main difference between these two simulations is the polarisation after single scattering $p$ (upper left panel of \pref{fig:fill_smax_850}), which decreases from \num{0.87} to \num{0.55} for the larger maximum grain size. At the same time, more of the larger grains are present in the disk, which tends to have lower equilibrium temperatures. Thus, the anisotropy of the radiation field is reduced as well. Both the lower temperature and the lower single-scattering polarisation lead to a reduced degree of polarisation for scattering only (upper left panel of \pref{fig:res_simu_mass_amax_por_q}) and to a reduced overall degree of polarisation (lower left panel of \pref{fig:res_simu_mass_amax_por_q}). A smaller maximum grain size, however, may lead to an increased net polarisation degree. Although the opacity, and thus the optical depth, is lower by about a factor \num{16} than the reference case for $s_{\text{max}} = \SI{100}{\um}$, the net polarisation is higher by a factor \num{1.15} due to the higher single-scattering polarisation, the steeper temperature distribution of the smaller grains, and the resulting high radiation field anisotropy. A maximum grain size of only $\SI{100}{\um}$ does not agree either with the commonly accepted dust and disk evolution models, however, nor do these models predict a sharp decrease of the grain size distribution beyond this value.

        Another possibility to increase the degree of polarisation might be to change the grain size distribution. In our reference model, the size distribution has a power-law exponent of $q=\num{-3.5}$. In \pref{sec:appendix_opt_prop}, the corresponding plots of the optical properties for different exponents can be found. The scattering polarisation steadily increases for steeper grain size distributions; see the upper right panel of \pref{fig:res_simu_mass_amax_por_q}. This is again due to the larger amount of smaller grains, which have a steeper temperature distribution and a higher single-scattering polarisation. The flux ratio of scattered to direct radiation, on the other hand, is only slightly altered. Thus, the overall degree of polarisation for compact and moderately porous grains increases for steeper grain size distributions as well. Nevertheless, the polarisation degree is still below the observed values.

\begin{figure}
            \includegraphics[width=0.49\linewidth]{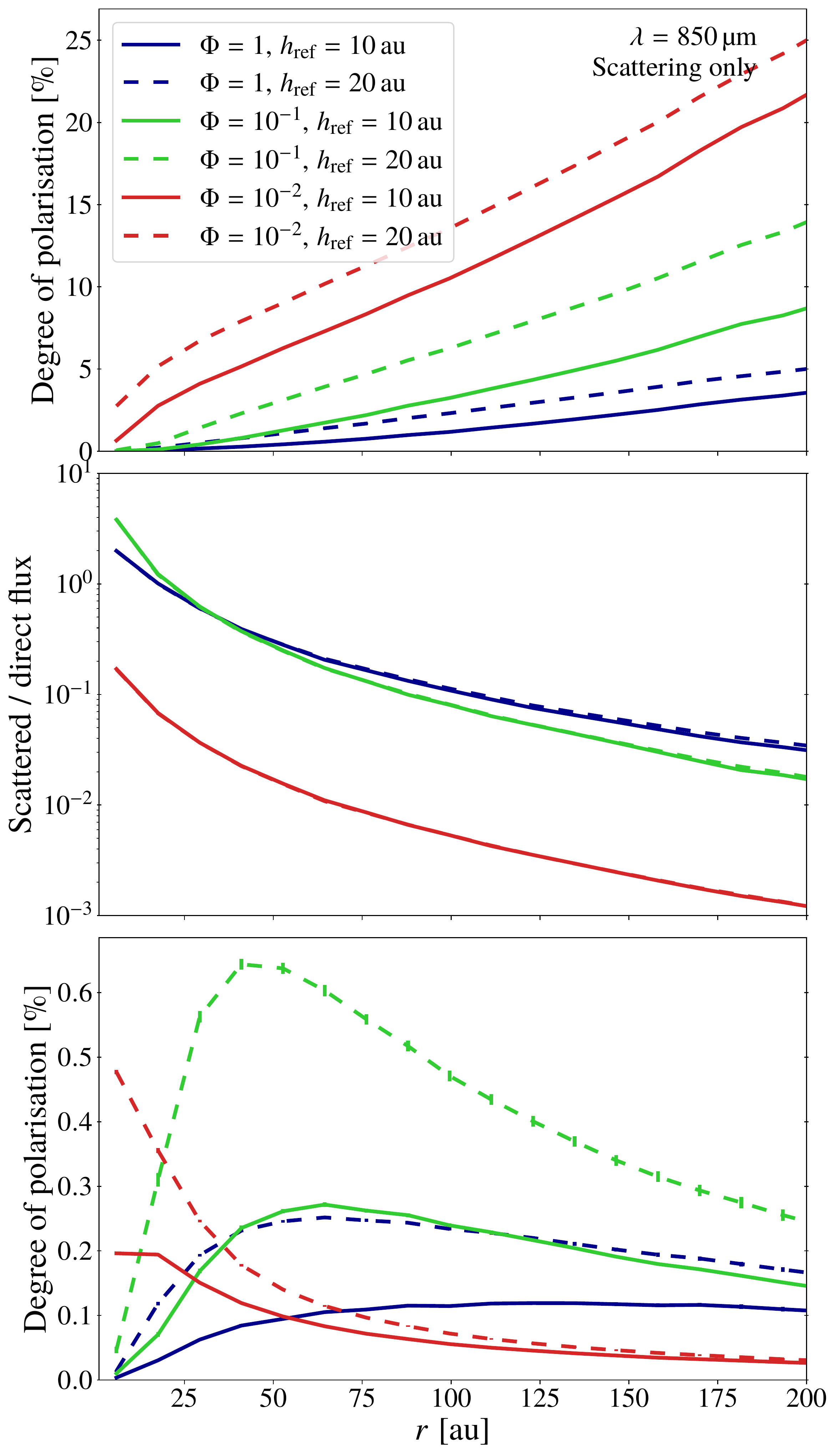}
            \includegraphics[width=0.49\linewidth]{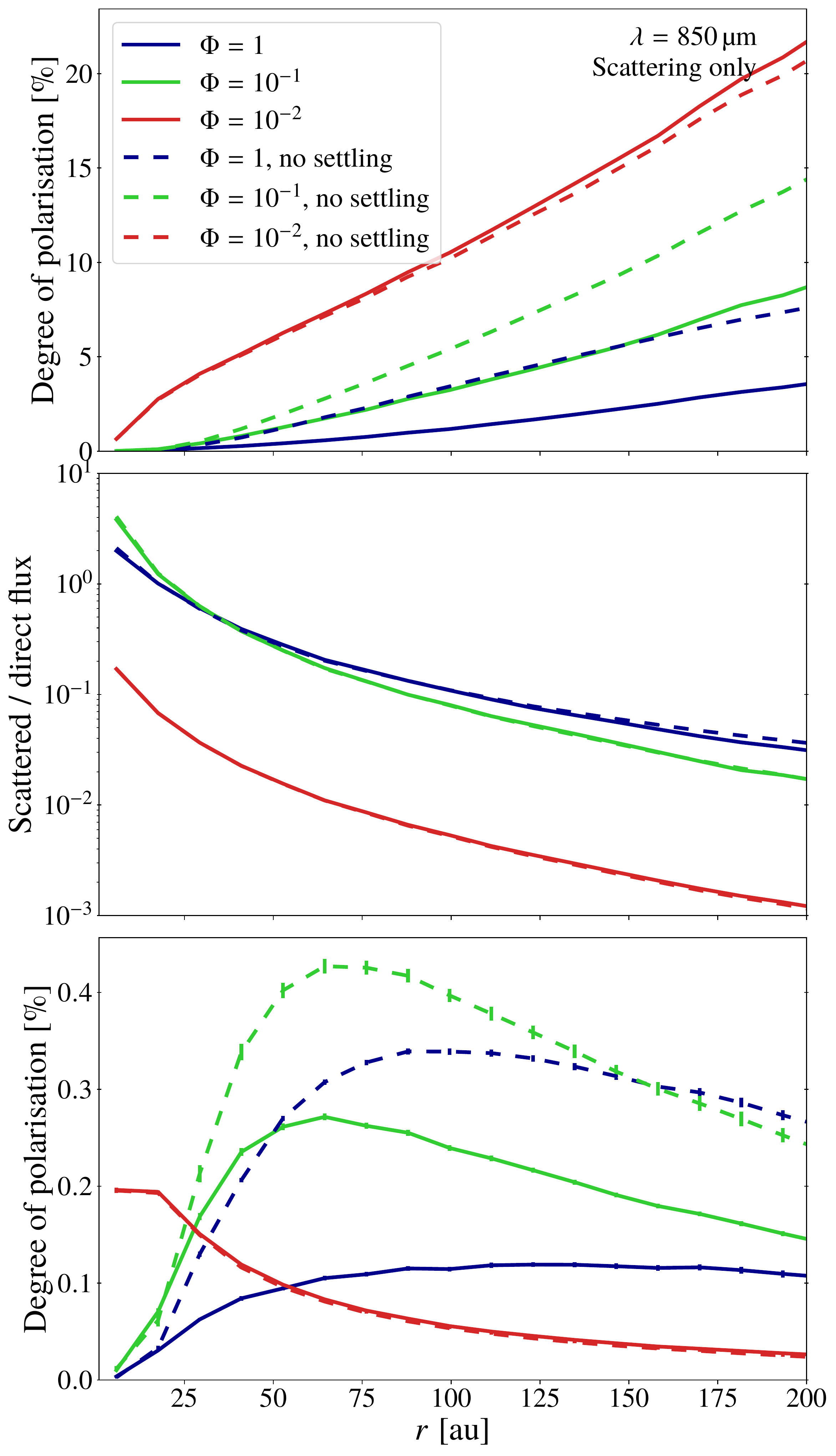}
            \caption{Same as \pref{fig:res_simu_mass_mass_amax_amax}, but for disks with different filling factors (colour) and reference scale heights (left), and without dust settling (right).}
            \label{fig:res_simu_por_h20_noSettling}
        \end{figure}

         \citet{brunngraeber-wolf-2020} showed that a larger scale height of the dust distribution increases the polarisation degree. This is even more pronounced for porous grains, as is shown in the left column of \pref{fig:res_simu_por_h20_noSettling}. The net polarisation rises above \SI{0.6}{\percent} for an intermediate filling factor $\Phi=\num{e-1}$ (green) and a scale height twice as large as the reference case $h_{\text{ref}}=\SI{20}{\au}$ (dashed). Although there is observational evidence for such large scale heights, most studies tend towards smaller scale heights of about $\SI{10}{\au}$ or smaller \citep{stapelfeldt-et-al-1998,andrews-et-al-2010,woitke-et-al-2019,villenave-et-al-2020}. When only the scale heights of the largest grains are increased by discarding the dust settling, the polarisation also increases up to about \SI{0.5}{\percent}; see the left column of \pref{fig:res_simu_por_h20_noSettling}. This is again contrary to the expected dust distribution from evolutionary models.

\begin{figure}
            \includegraphics[width=0.49\linewidth]{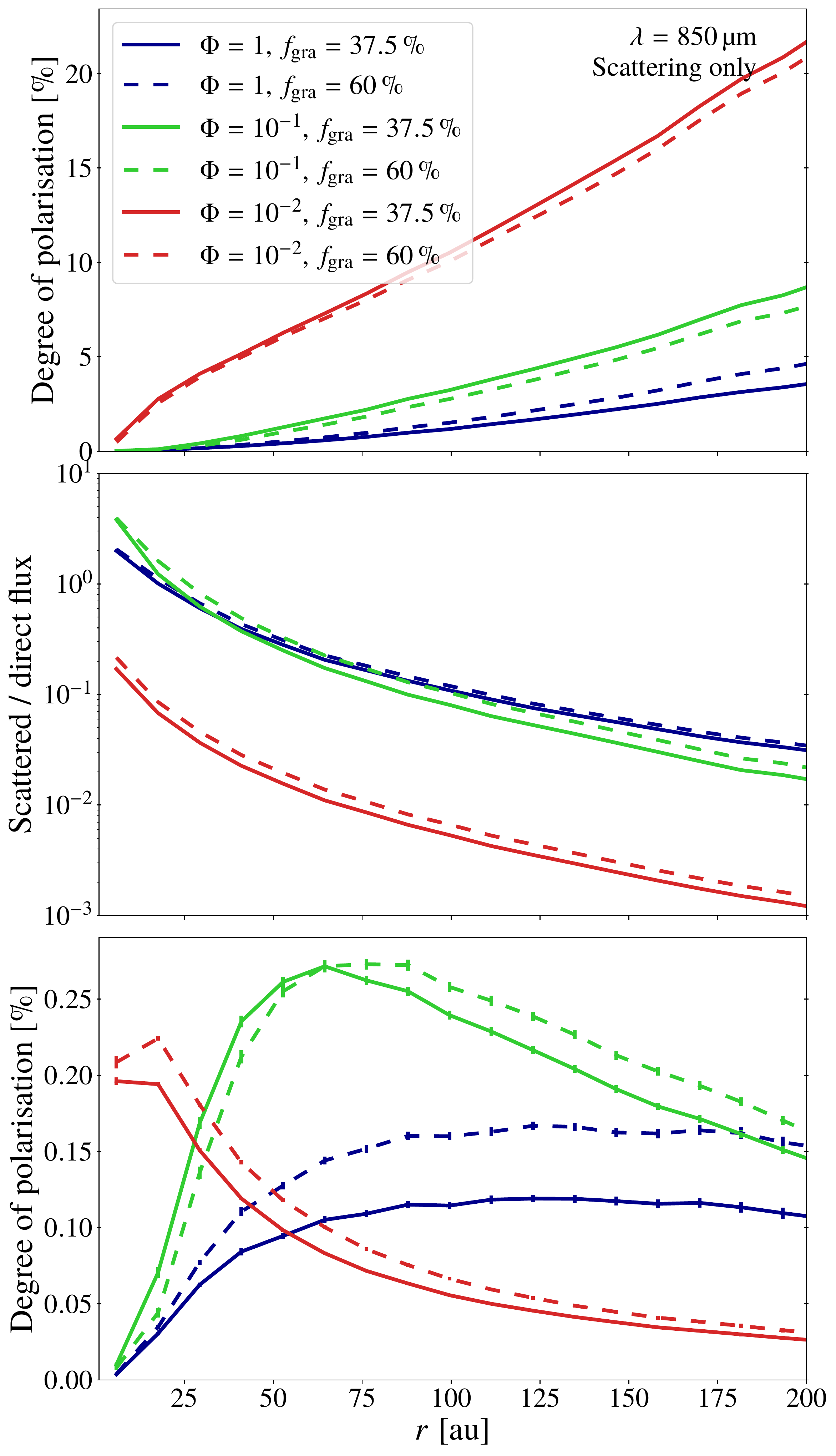}
            \caption{Same as \pref{fig:res_simu_mass_mass_amax_amax}, but for disks with different filling factors (colours) and mass fraction of graphite in the dust model (line style).}
            \label{fig:res_simu_por_sili}
        \end{figure}

        Driven by the goal to achieve a higher opacity and single-scattering polarisation and thus to reproduce the generally observed polarisation degree in the order of a few percent, a high amount of carbonaceous dust was considered by \citet{yang-li-2020}. We partially confirm their findings. When we increased the fraction of graphite in our dust model from \SI{37.5}{\percent} to \SI{60}{\percent}, the polarisation indeed increased, but only by a small amount, and the polarisation is still well below \SI{1}{\percent}. The maximum value is about \SI{0.3}{\percent}; see \pref{fig:res_simu_por_sili}. This means that a large graphite fraction alone does not sufficiently increase the polarisation degree.

\begin{figure}
            \includegraphics[width=0.95\linewidth]{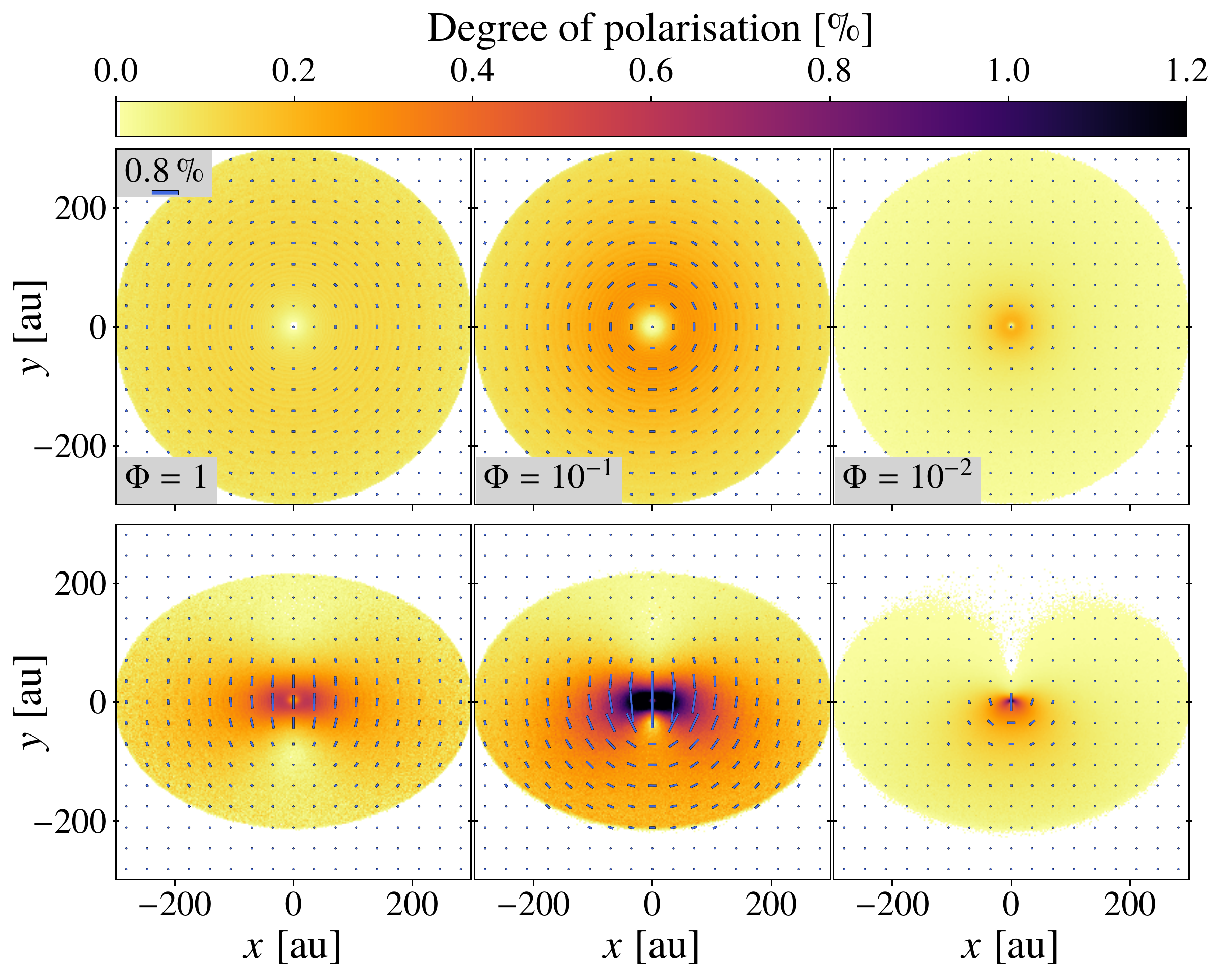}
            \caption{Map of the polarisation degree with superimposed polarisation vectors for our reference disk with $\Phi=1$ (\textit{left}), \num{e-1} (\textit{centre}), and \num{e-2} (\textit{right}), and an inclination of $i=\SI{0}{\degree}$ (\textit{top}) and $\SI{45}{\degree}$ (\textit{bottom}). The lengths of the polarisation vectors are scaled to the polarisation degree, as indicated in the top left corner.}
            \label{fig:pol_map_reference}
        \end{figure}

\begin{figure}
            \includegraphics[width=0.95\linewidth]{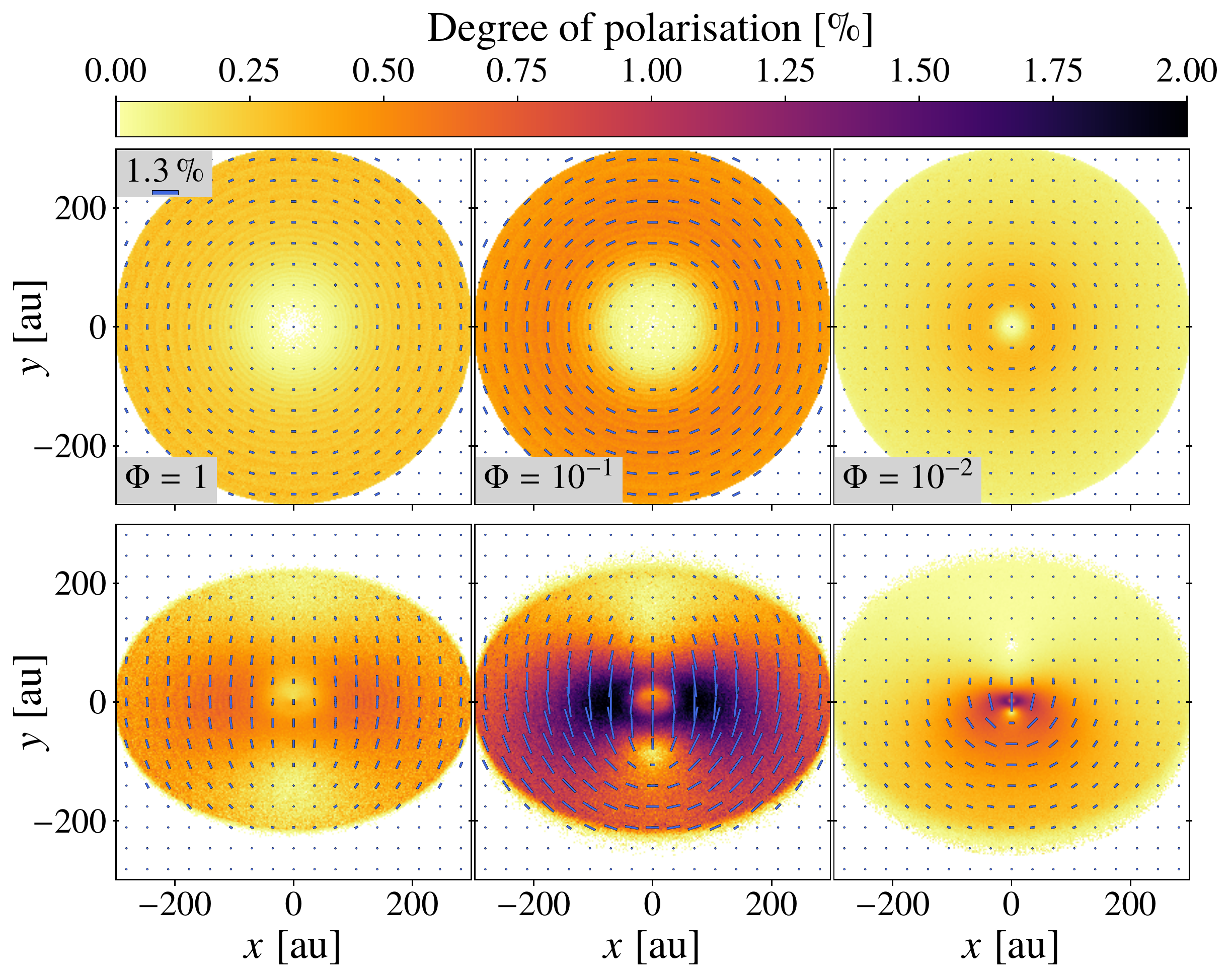}
            \caption{Same as \pref{fig:pol_map_reference}, but for a disk with a ten times higher dust mass. The lengths of the polarisation vectors are scaled to the polarisation degree, as indicated in the top left corner.}
            \label{fig:pol_map_m1e-3}
        \end{figure}

        As already shown in several studies \citep[e.g.][]{yang-et-al-2016a,yang-et-al-2017}, the polarisation degree also depends on the disk inclination. We confirm these findings as shown in \pref{fig:pol_map_reference} with our reference model and an inclination of \SI{45}{\degree}. Whereas the polarisation of scattered flux only is very similar for both viewing angles, the ratio of the scattered to direct flux increases especially at the near side of the disk, and hence the increased net polarisation fraction up to \SI{1.9}{\percent} in the innermost regions for a filling factor of $\num{0.1}$ (middle panel). This inclination-induced increase of the polarisation arises because the scattering angle and the resulting polarisation is different for radiation parallel to the major disk axis than for radiation parallel to the minor disk axis. Thus, the superposition of these contributions is less destructive and hence results in a lower decrease in polarisation degree than in the face-on case, where the scattering angles towards the observer at a given point in the disk are very similar for radiation from all directions. This is described in detail in \citet{yang-et-al-2016a}. Thus, with an inclined disk consisting of moderately porous mm-sized grains, we are able to achieve levels of polarisation that are comparable to selected observations \citep{bacciotti-et-al-2018,lee-et-al-2018,mori-et-al-2019,ohashi-kataoka-2019}. These trends concerning filling factor, size distribution, and density distribution still hold for inclined disks. Additionally, we stress that the disk mass has a significant effect on the level and spatial distribution of the polarisation degree for inclined disks through the optical depth. The optical depth and the mass under the assumption of a known dust model can therefore in principle be constrained by spatially resolved polarisation observations; see \pref{fig:pol_map_m1e-3}.

\section{Discussion}
\label{sec:discussion}
    The degree of polarisation resulting from our simulations is in most cases lower by a factor 2 to 3 than what is observed in several protoplanetary disks where the polarisation is thought to result from scattering. Although both albedo $\omega$ and single-scattering polarisation $p=-\nicefrac{S_{12}}{S_{11}}$ imply a high degree of polarisation for decreasing filling factors $\Phi$, the net degree of polarisation of the total radiation increases only for a small range of filling factors before decreasing towards even lower filling factors. This decrease in polarisation after reaching a maximum value at about $\Phi=\num{0.1}$ is due to the decrease in extinction opacity of the dust grains and the resulting low optical depth, and hence a decrease in the scattering probability.

    This might in principle be addressed by different modifications of the model. To increase the optical depth and thus the scattering probability, intuitive steps would be to increase the disk mass, the maximum grain size, or use a more shallow grain size distribution. However, these changes do not necessarily lead to higher polarisation degrees, as seen in the previous section. \citet{yang-li-2020} showed that a high abundance of carbonaceous material would increase the single-scattering polarisation as well. \pref{sec:appendix_opt_prop} shows that the opacity and albedo are indeed higher than in astrosil grains. However, our reference dust model already contains \SI{37.5}{\percent} graphite (mass fraction). Even a larger fraction of \SI{60}{\percent} graphite increases the polarisation only by a small amount, the maximum degree of polarisation is still below \SI{0.3}{\percent} for a face-on disk.

    Although the degree of polarisation significantly increases for inclined disks, by a factor of about 4 for our reference disk and an inclination of \SI{45}{\degree} , it remains below most of the observed polarisation degrees when compact grains are assumed. For a moderate filling factor of \num{0.1}, our simulations reach polarisation degrees of $\SI{1.9}{\percent}$ in the innermost disk regions, which is in agreement with several observations. An only moderately porous dust phase is also in agreement with recent studies concerning the interstellar medium \citep{hirashita-et-al-2021,draine-hensley-2021}, and the rotational disruption of dust grains in protoplanetary disks \citep{tatsuuma-kataoka-2021}. Furthermore, the polarisation of inclined disks follows the same trends concerning the filling factor as mentioned above in the face-on case.

    Nevertheless, the spectral index of our simulated disks between the wavelengths $\lambda=\SI{1.3}{\mm}$ and $\SI{2}{\mm}$ significantly increases for porous grains compared to compact grains. This is in contrast to many observations of protoplanetary disks where the spectral index at \si{mm} wavelengths is found to be in the range between \num{2} and \num{2.5} which is commonly interpreted as evidence for large grains \citep{andrews-williams-2005}. In our simulations, however, the spectral index of disks consisting of porous grains is larger than \num{3} and is as high as \num{3.8} for low-mass disks. Thus, our results suggest that the grain porosity $\mathcal{P}=1-\Phi$ in protoplanetary disks does not exceed $0.9$ significantly and may be even lower. Although this is in agreement with \citet{tazaki-et-al-2019}, for instance, it cannot be ruled out that larger dust grains of about several cm in radius are present in the disk and depress the spectral index. At the same time, however, this would decrease the polarisation fraction.

    A common suggestion for increasing the scattering polarisation in disk models without the inherent problem with the spectral index in the case of porous grains outlined above is introducing non-spherical dust grains. \citet{kirchschlager-bertrang-2020} showed that non-spherical compact, that is, non-porous, silicate dust grains might show a higher single-scattering polarisation degree. The complexity of the dust model increases significantly with non-spherical dust grains, and the prediction of whether the observable degree of polarisation would increase or decrease is not straightforward. This uncertainty is even larger as the alignment efficiency and the dominating alignment mechanism in protoplanetary disks is still unclear. In addition to the polarisation degree, the observed orientations place further constraints on the origin of the polarisation and need to be accounted for as well. A proper investigation including extensive Monte Carlo radiative transfer simulations remains to be done.

\section{Summary}
\label{sec:summary}
    We investigated the effect of dust grain porosity on the polarisation degree due to scattering of thermal re-emission radiation in the sub-mm wavelength range. For this purpose, we analysed the optical properties of grain size distributions with different filling factors and upper grain size limits. This analysis approach is independent of any underlying density distribution and thus independent of the local radiation field, and allows general discussions of the polarisation.

    We showed that focussing on the single-scattering polarisation $p=-\nicefrac{S_{12}}{S_{11}}$ and the albedo $\omega$ is not sufficient for reliable estimates of the overall degree of polarisation. We stress that the extinction opacity and the flux distribution for different scattering angles have to be included in these considerations. Especially for larger grains, the scattered flux decreases dramatically for scattering angles close to $\SI{90}{\degree}$ as large grains tend to show a strong forward-scattering behaviour. Furthermore, increasing the upper grain size limit does not necessarily increase the opacity or optical depth of the entire dust phase. Although the extinction cross-section is approximately proportional to the grain surface and thus to $s^2$, the actual number of dust grains is indirectly proportional to the grain mass, that is, to $s^{-3}$, resulting in a decrease in opacity.

    In the second part of our study, full radiative transfer simulations of typical protoplanetary disks with porous dust grains were conducted. Here, the largest grains are already settled towards the midplane to account for dust settling. It was shown that introducing porosity may increase the degree of polarisation by up to a factor of four. However, decreasing the filling factor below $\sim \num{0.1}$ will decrease the degree of polarisation for even lower filling factors because of the low opacity and optical depth. The flux of the unpolarised direct emission of the dust becomes much higher than the scattered flux, which effectively decreases the polarisation. Furthermore, the spectral index of disks consisting of porous grains is significantly larger than observed.

    In the face-on case, increasing the anisotropy of the radiation field by larger temperature differences in the disk may lead to polarisation degrees in the order of what is observed in protoplanetary disks. This can be achieved for instance by rather small upper grain size limits in the order of \SI{100}{\um} or a significant overabundance of grains in that size range, or by a high anisotropy of the radiation field due to rather extreme density distributions, caused by large scale heights, for example. Both are very unlikely scenarios for typical protoplanetary disks according to current disk and dust evolution models.

    For inclined disks, our simulations are able to reproduce polarisation degrees of \SI{1.2}{\percent} and higher with our reference disk model including \si{mm}-sized grains, moderate porosity, dust settling, and reasonable (i.e. in agreement with accepted evolution theories and observations) geometrical disk properties. However, the spectral index of these simulations does not reproduce the observed values in the \si{mm} wavelength range.

    The mismatch between the spatially resolved polarimetric observations at sub-mm wavelengths, the observed spectral index and the predicted values resulting from theoretical models, and the interpretation of a wealth of previous observations of dust in protoplanetary disks remains an open issue that urgently needs to be further addressed in future studies.

\begin{acknowledgement}
    This research was funded through the DFG grant WO 857/18-1. This research made use of Astropy\footnote{http://www.astropy.org}, a community-developed core Python package for Astronomy \citep{astropy-2018}.
\end{acknowledgement}




\bibliographystyle{aa}
\bibliography{lit}



\begin{appendix}
    \section{Optical properties for further dust mixtures}
    \label{sec:appendix_opt_prop}
        \begin{figure*}
            \includegraphics[width=0.49\linewidth, height=0.18\textheight]{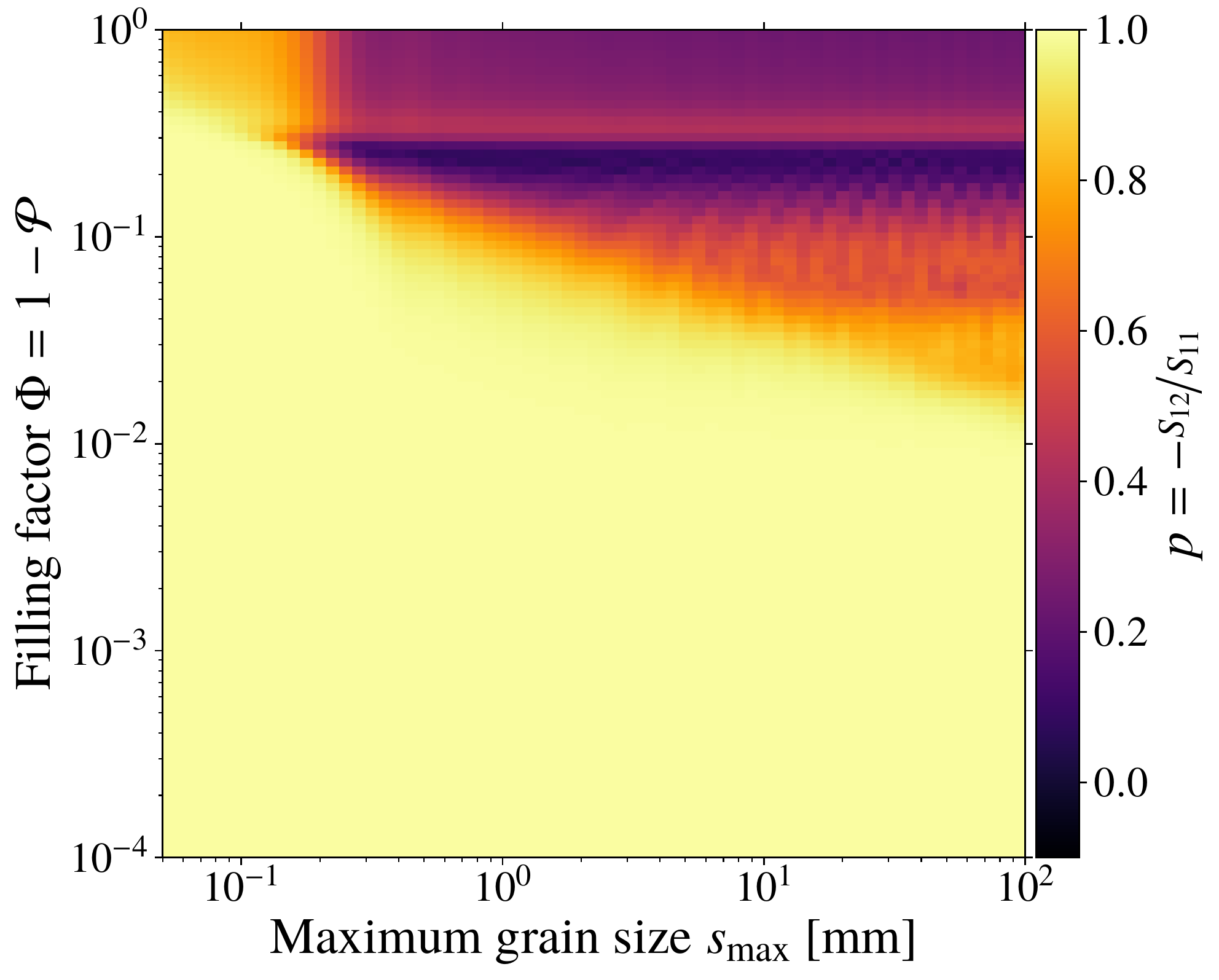}
            \includegraphics[width=0.49\linewidth, height=0.18\textheight]{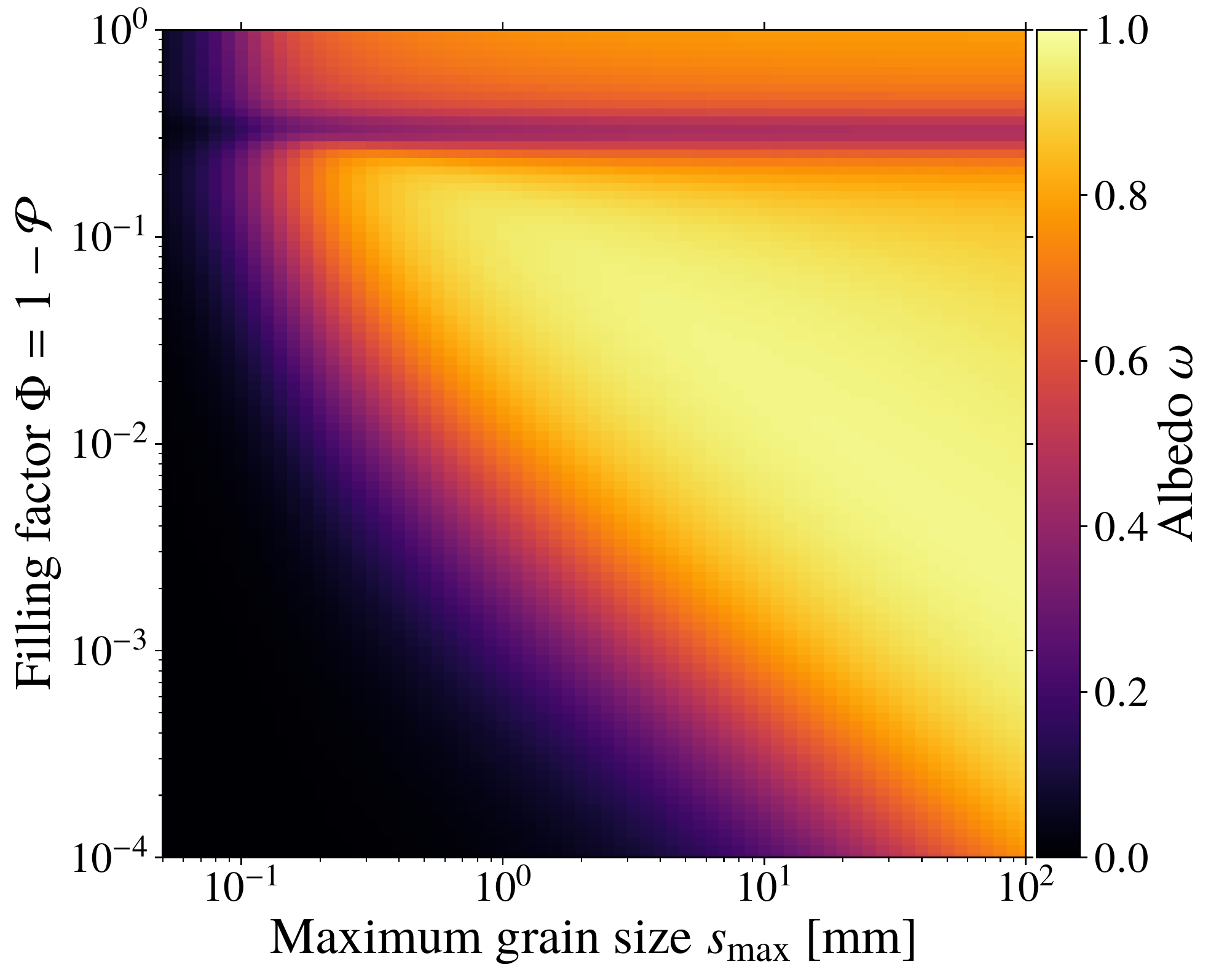}\\
            \includegraphics[width=0.49\linewidth, height=0.18\textheight]{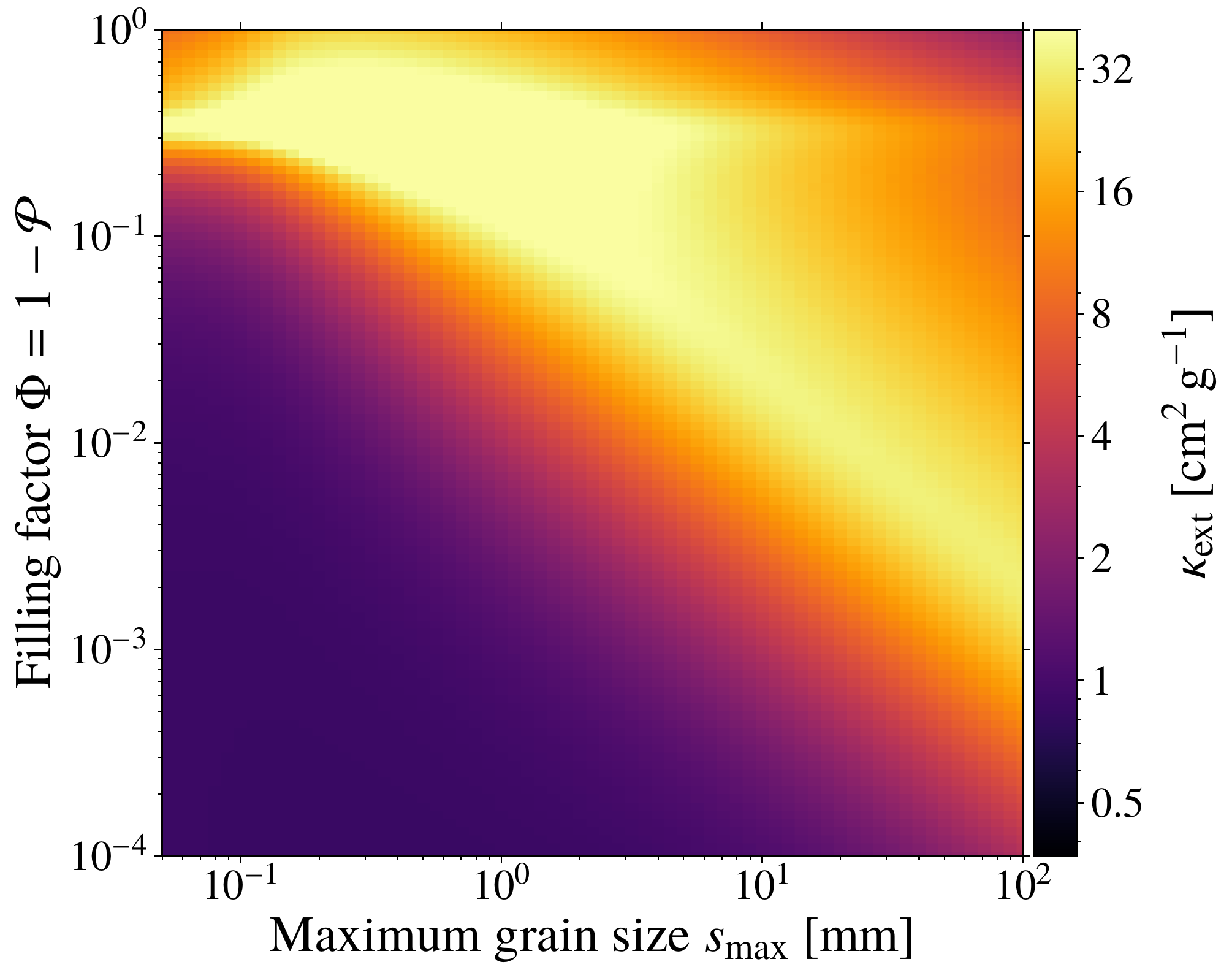}
            \includegraphics[width=0.49\linewidth, height=0.18\textheight]{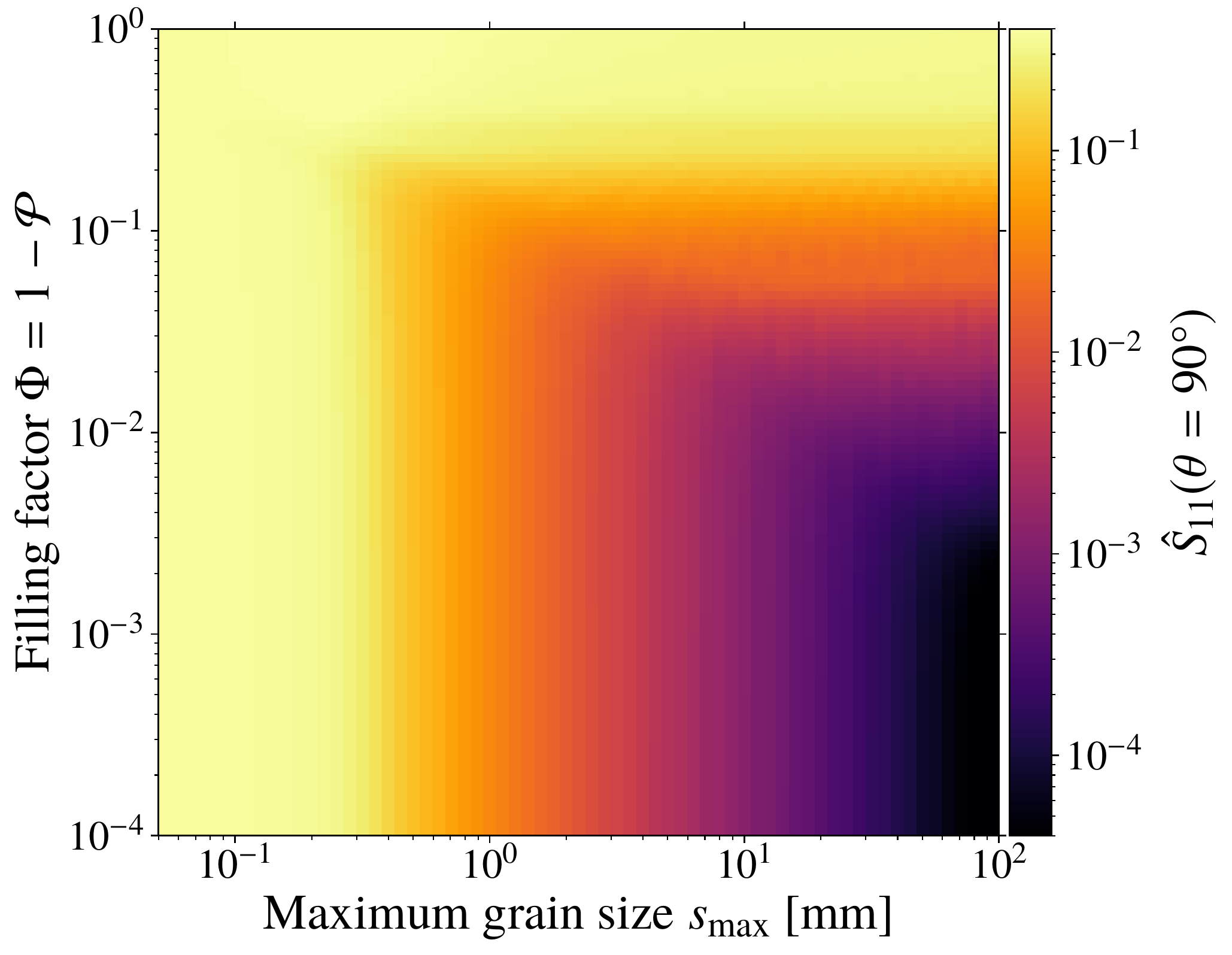}
            \caption{Degree of polarisation for single scattering $p = -\nicefrac{S_{12}}{S_{11}}$ for a scattering angle of $\theta=\SI{90}{\degree}$ \textit{(upper left)}, albedo \textit{(upper right)}, extinction opacity \textit{(lower left)}, and probability density function $\hat{S}_{11}$ for the scattering angle $\theta=\SI{90}{\degree}$ \textit{(lower right)}. All quantities were calculated for a grain size distribution with $q=\num{-3.5}$, $s_{\text{min}}=\SI{5}{\nm}$, and different upper grain size limits $s_{\text{max}}$ at a wavelength of \SI{850}{\um} for graphite (parallel).}
            \label{fig:fill_smax_850_grapar}
        \end{figure*}
        \begin{figure*}
            \includegraphics[width=0.49\linewidth, height=0.18\textheight]{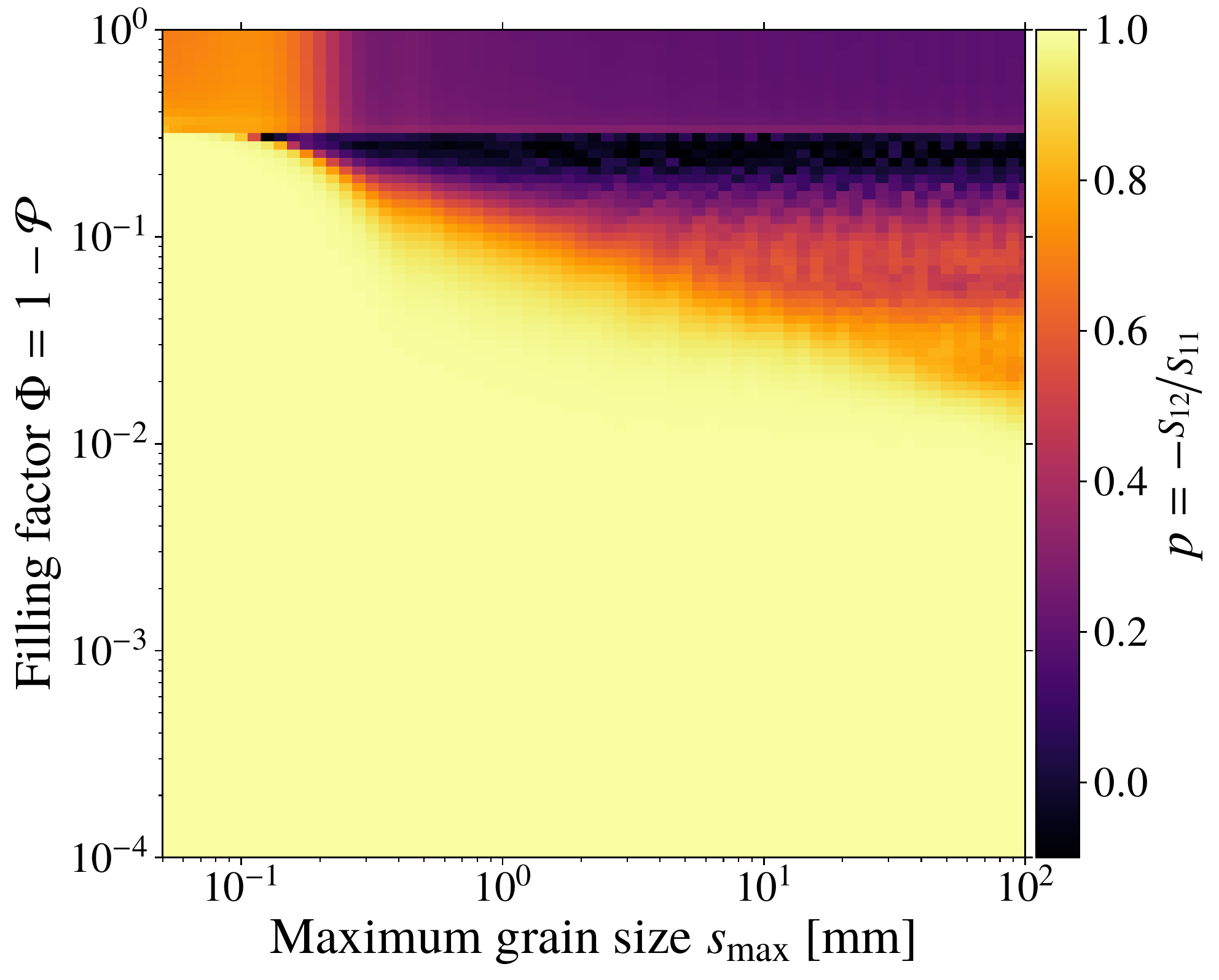}
            \includegraphics[width=0.49\linewidth, height=0.18\textheight]{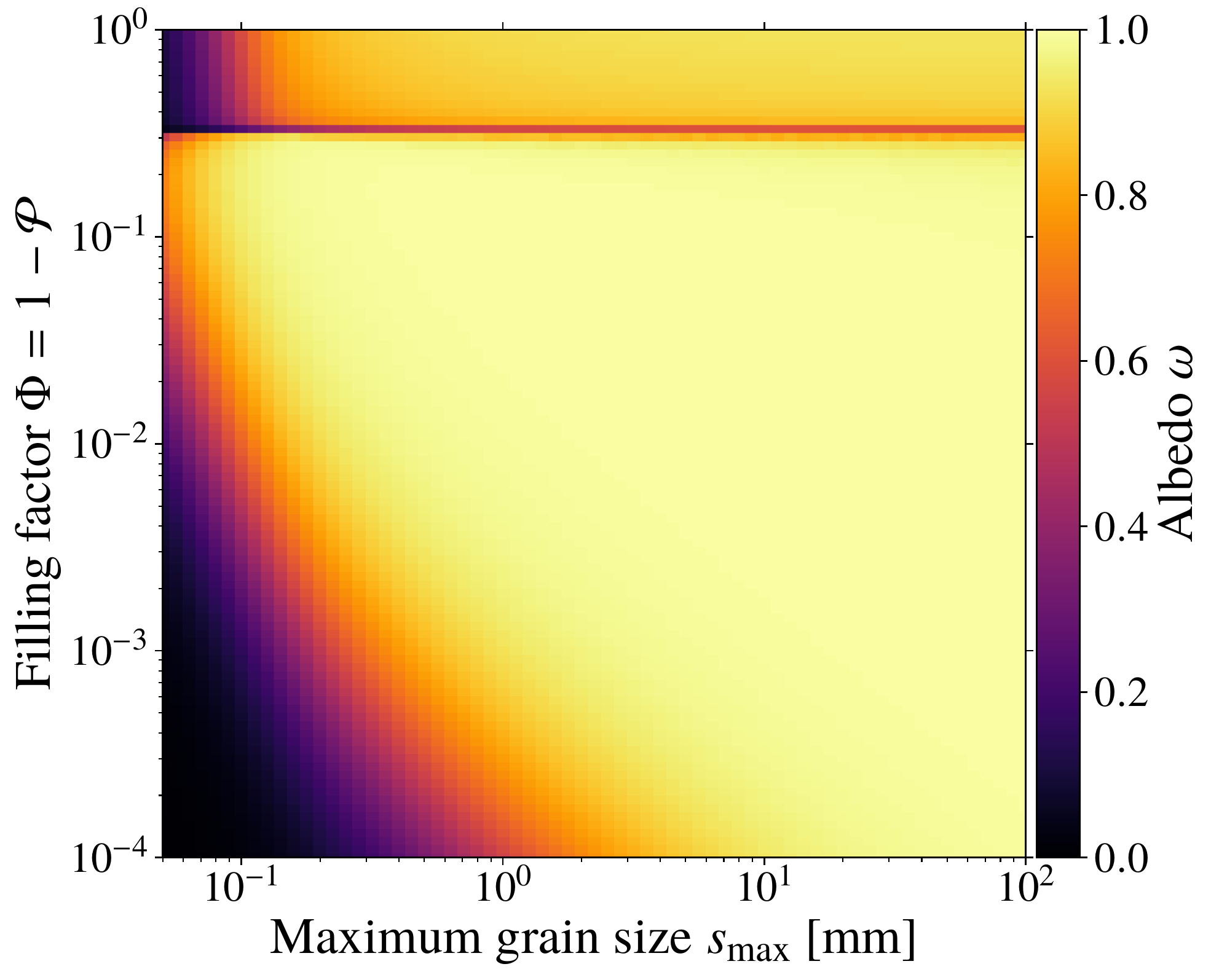}\\
            \includegraphics[width=0.49\linewidth, height=0.18\textheight]{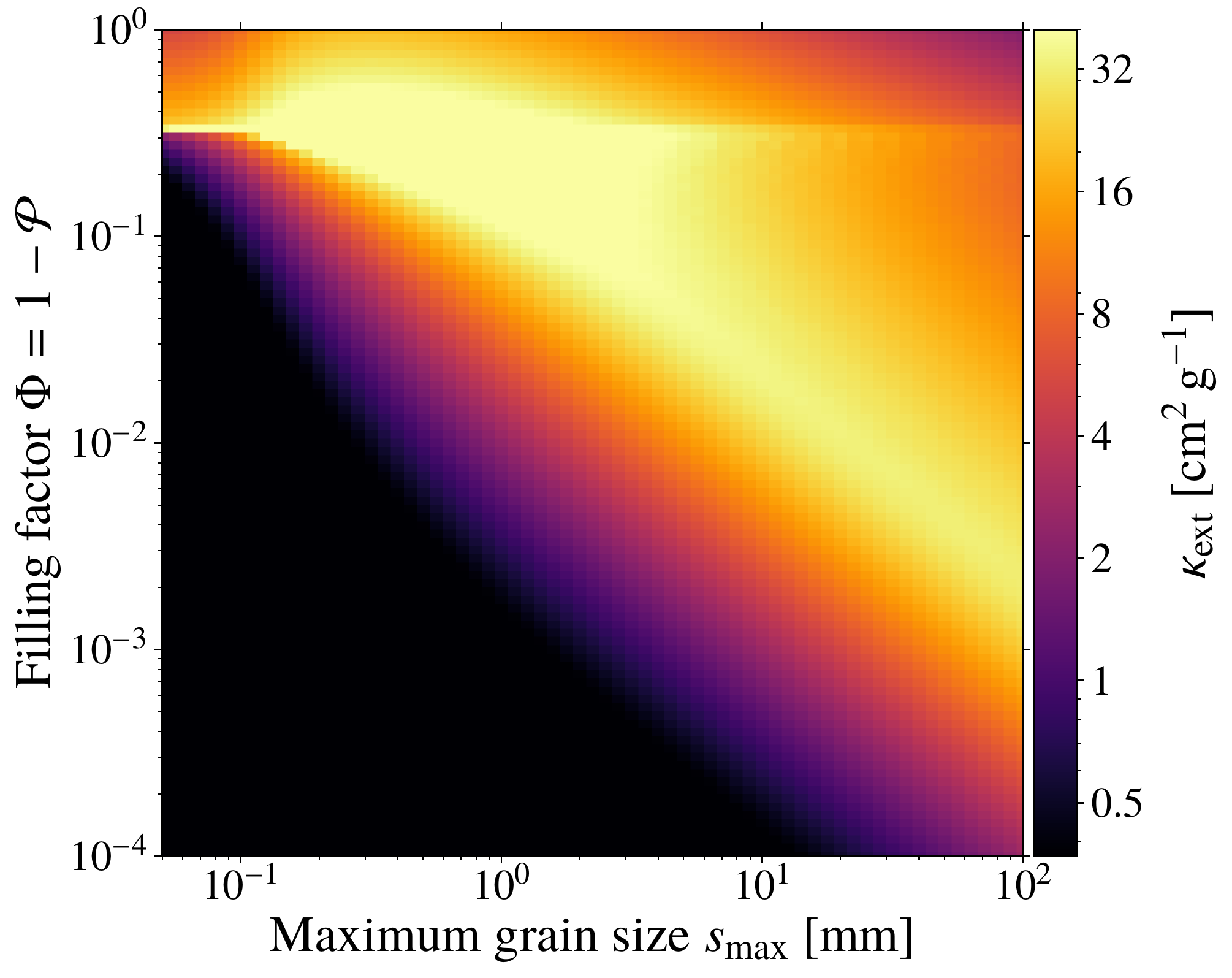}
            \includegraphics[width=0.49\linewidth, height=0.18\textheight]{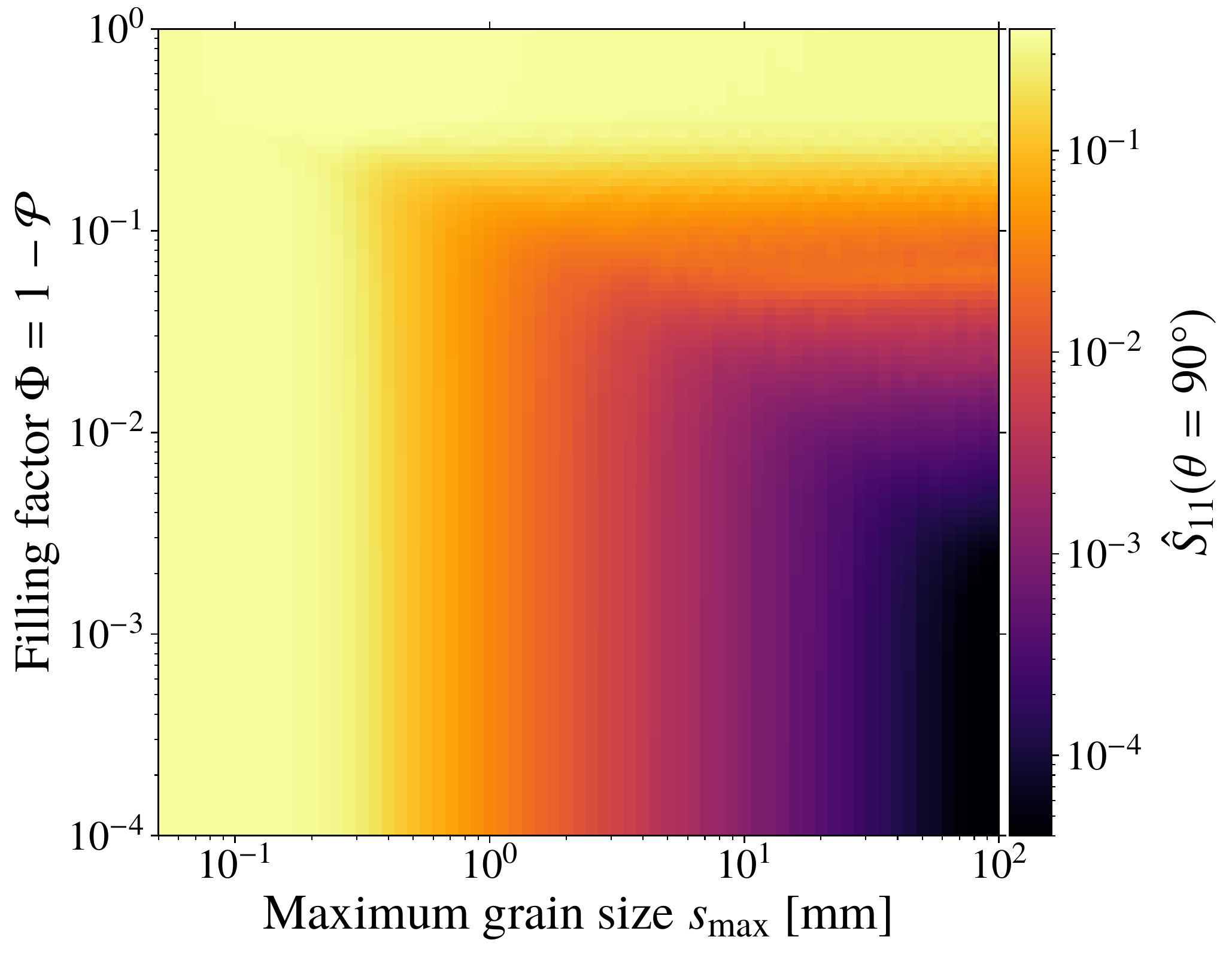}
            \caption{Degree of polarisation for single scattering $p = -\nicefrac{S_{12}}{S_{11}}$ for a scattering angle of $\theta=\SI{90}{\degree}$ \textit{(upper left)}, albedo \textit{(upper right)}, extinction opacity \textit{(lower left)}, and probability density function $\hat{S}_{11}$ for the scattering angle $\theta=\SI{90}{\degree}$ \textit{(lower right)}. All quantities were calculated for a grain size distribution with $q=\num{-3.5}$, $s_{\text{min}}=\SI{5}{\nm}$, and different upper grain size limits $s_{\text{max}}$ at a wavelength of \SI{850}{\um} for graphite (perpendicular).}
            \label{fig:fill_smax_850_graper}
        \end{figure*}

        \begin{figure*}
            \includegraphics[width=0.49\linewidth, height=0.18\textheight]{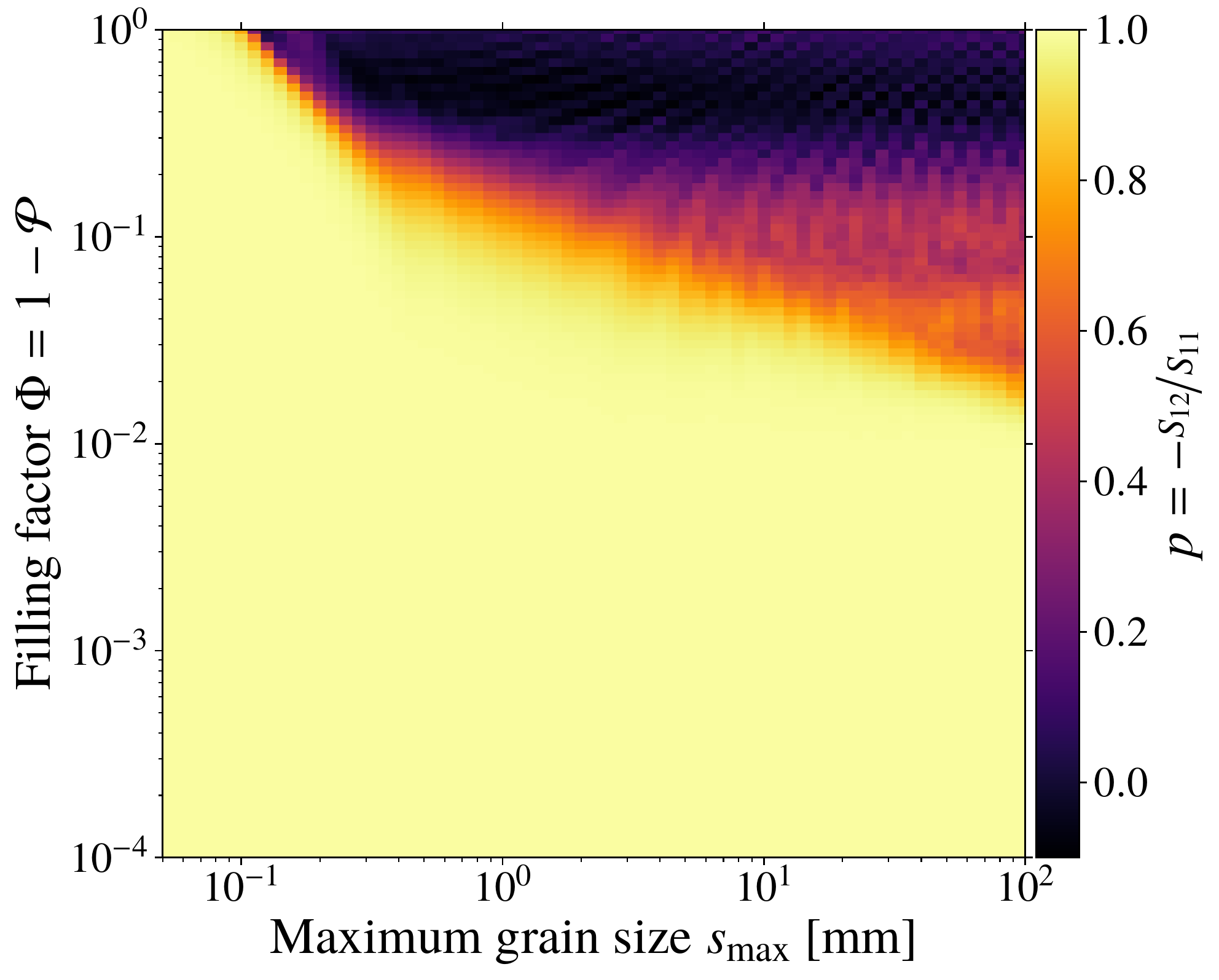}
            \includegraphics[width=0.49\linewidth, height=0.18\textheight]{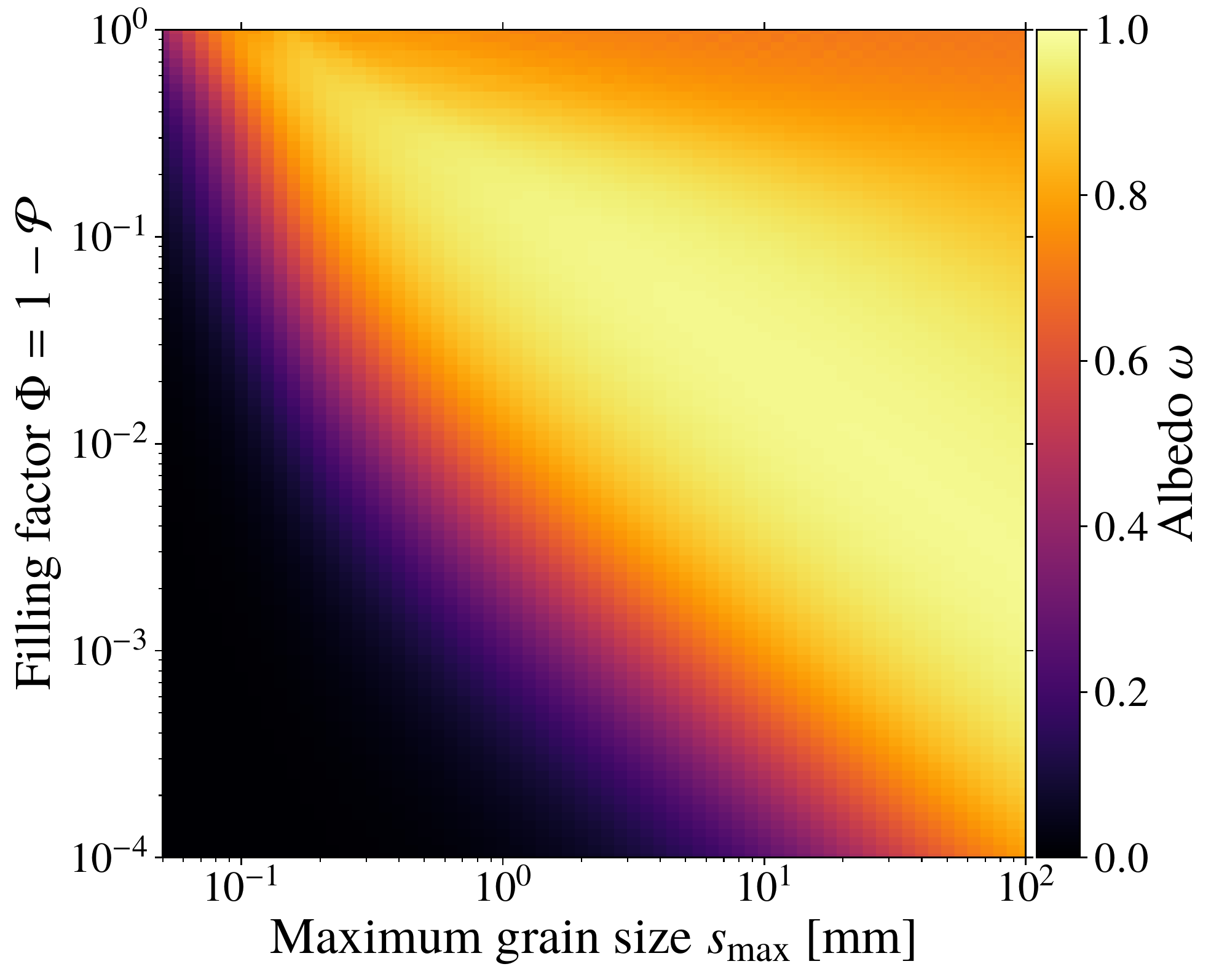}\\
            \includegraphics[width=0.49\linewidth, height=0.18\textheight]{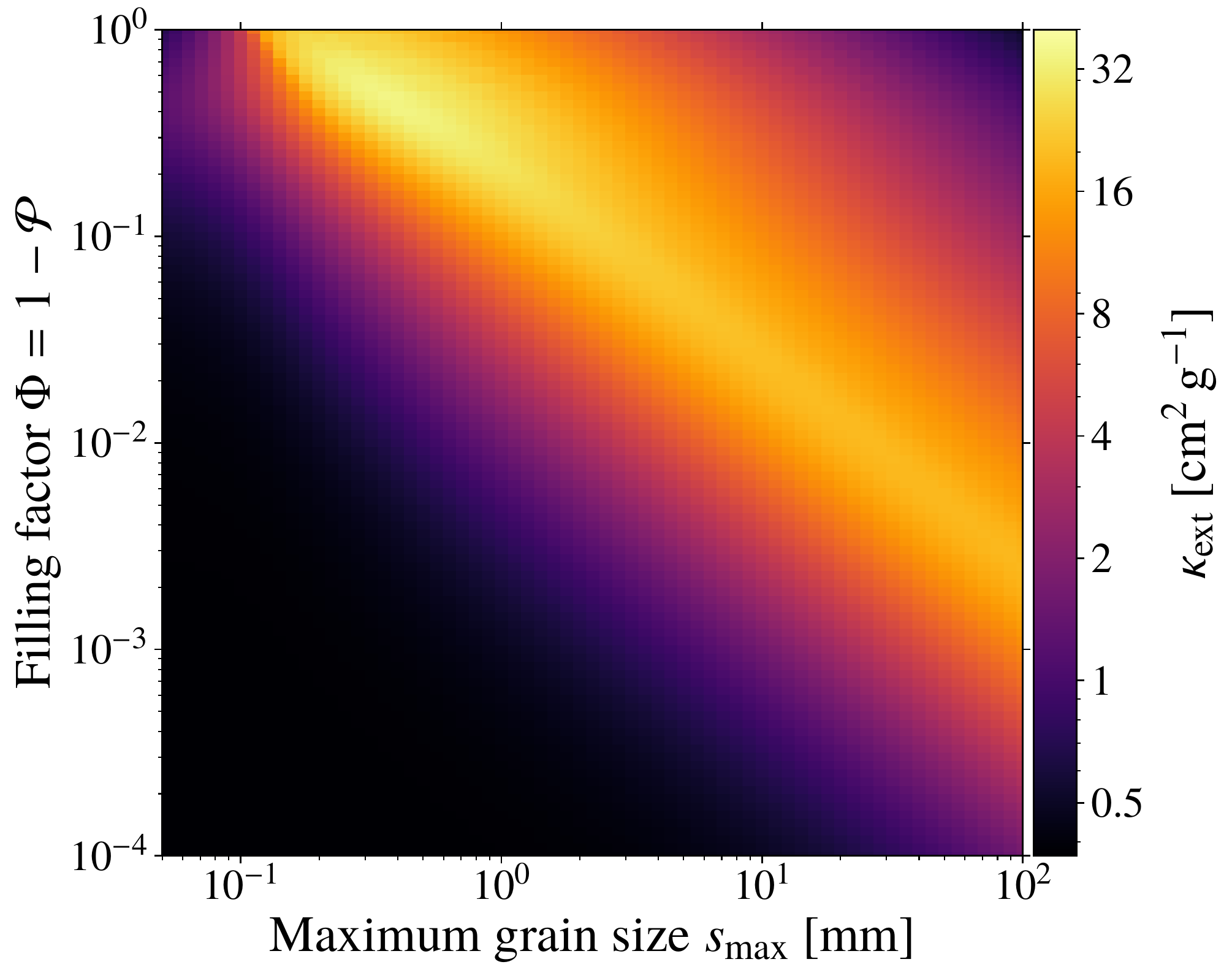}
            \includegraphics[width=0.49\linewidth, height=0.18\textheight]{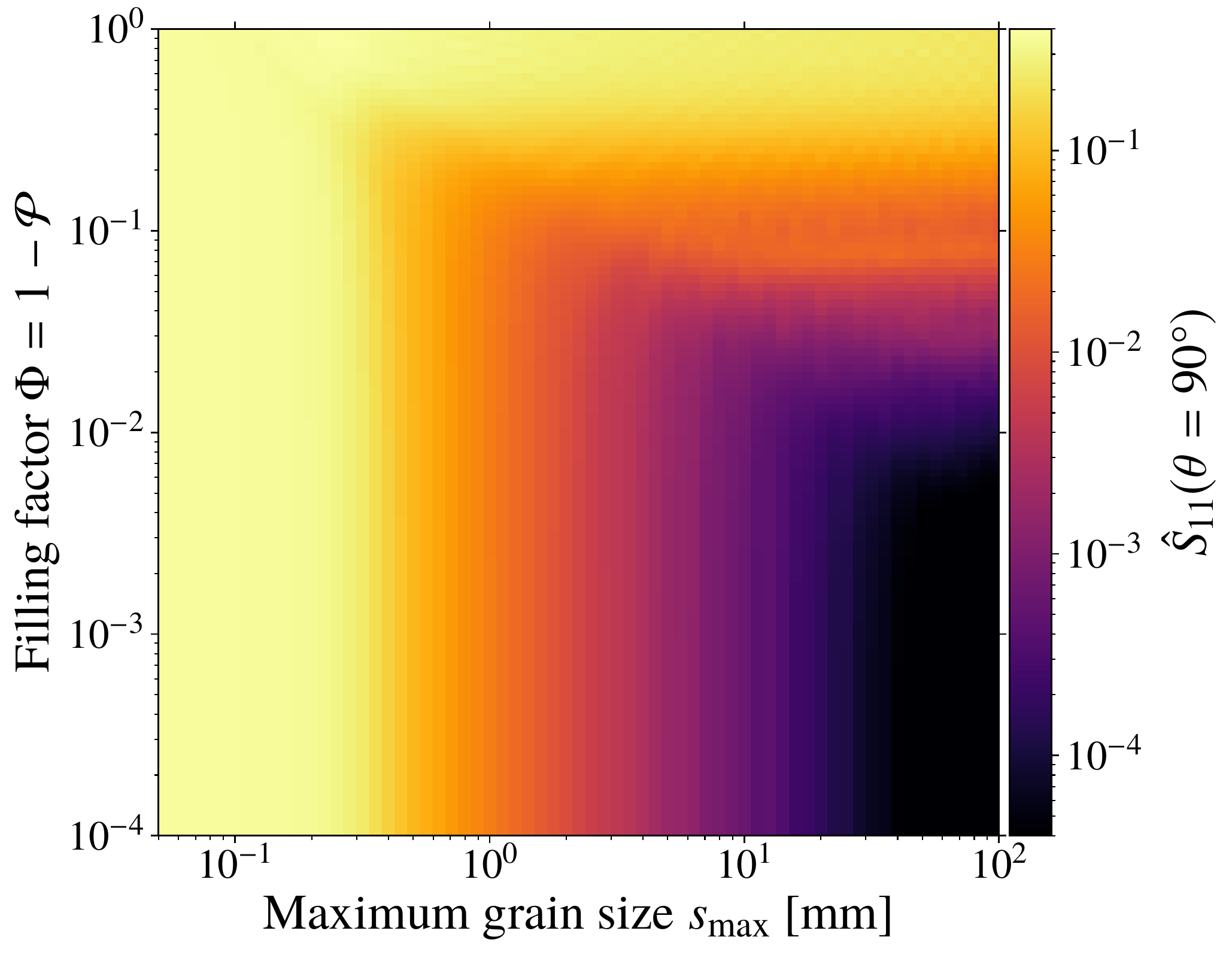}
            \caption{Degree of polarisation for single scattering $p = -\nicefrac{S_{12}}{S_{11}}$ for a scattering angle of $\theta=\SI{90}{\degree}$ \textit{(upper left)}, albedo \textit{(upper right)}, extinction opacity \textit{(lower left)}, and probability density function $\hat{S}_{11}$ for the scattering angle $\theta=\SI{90}{\degree}$ \textit{(lower right)}. All quantities were calculated for a grain size distribution with $q=\num{-3.2}$, $s_{\text{min}}=\SI{5}{\nm}$, and different upper grain size limits $s_{\text{max}}$ at a wavelength of \SI{850}{\um} for astrosil.}
            \label{fig:fill_smax_850_32}
        \end{figure*}
        \begin{figure*}
            \includegraphics[width=0.49\linewidth, height=0.18\textheight]{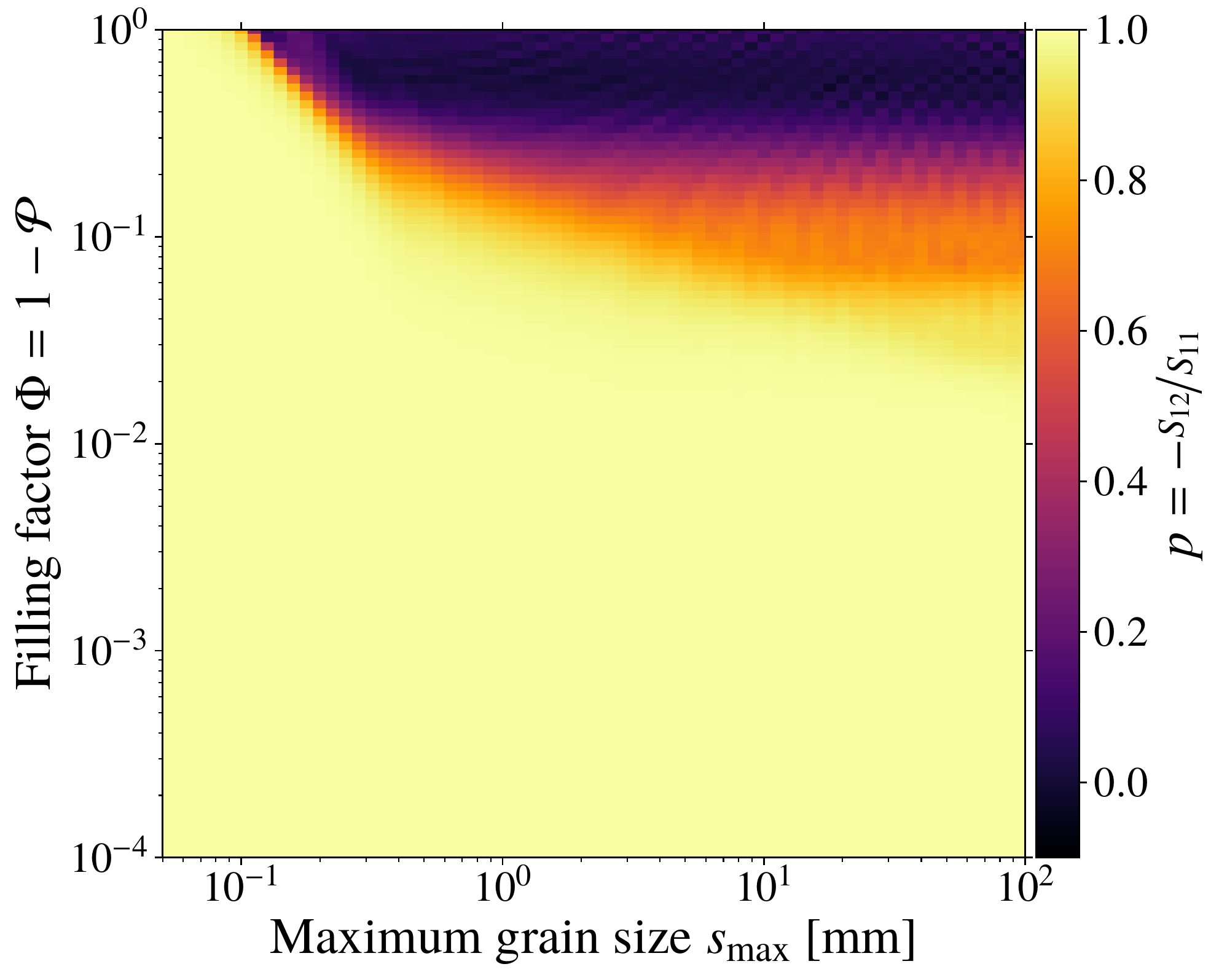}
            \includegraphics[width=0.49\linewidth, height=0.18\textheight]{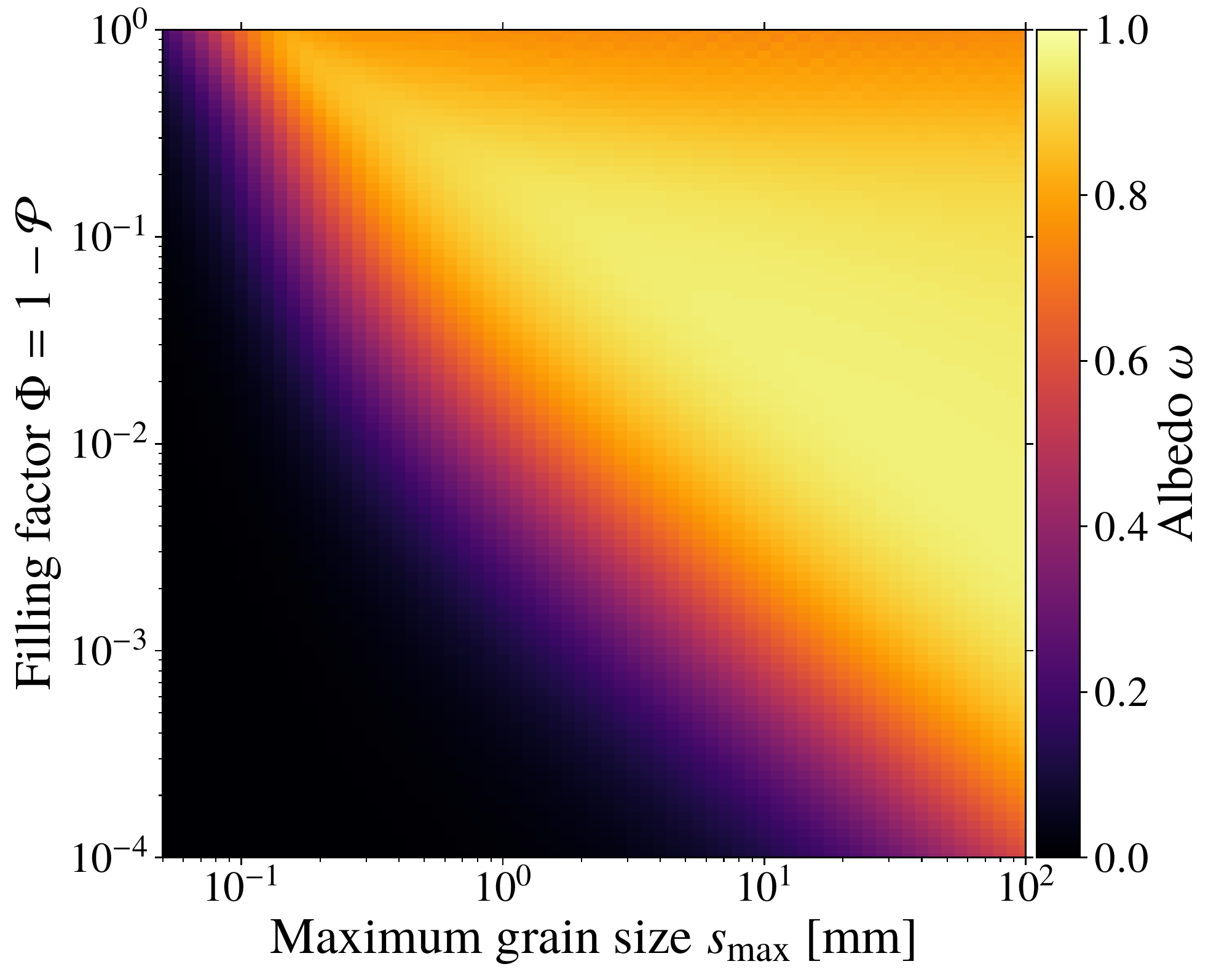}\\
            \includegraphics[width=0.49\linewidth, height=0.18\textheight]{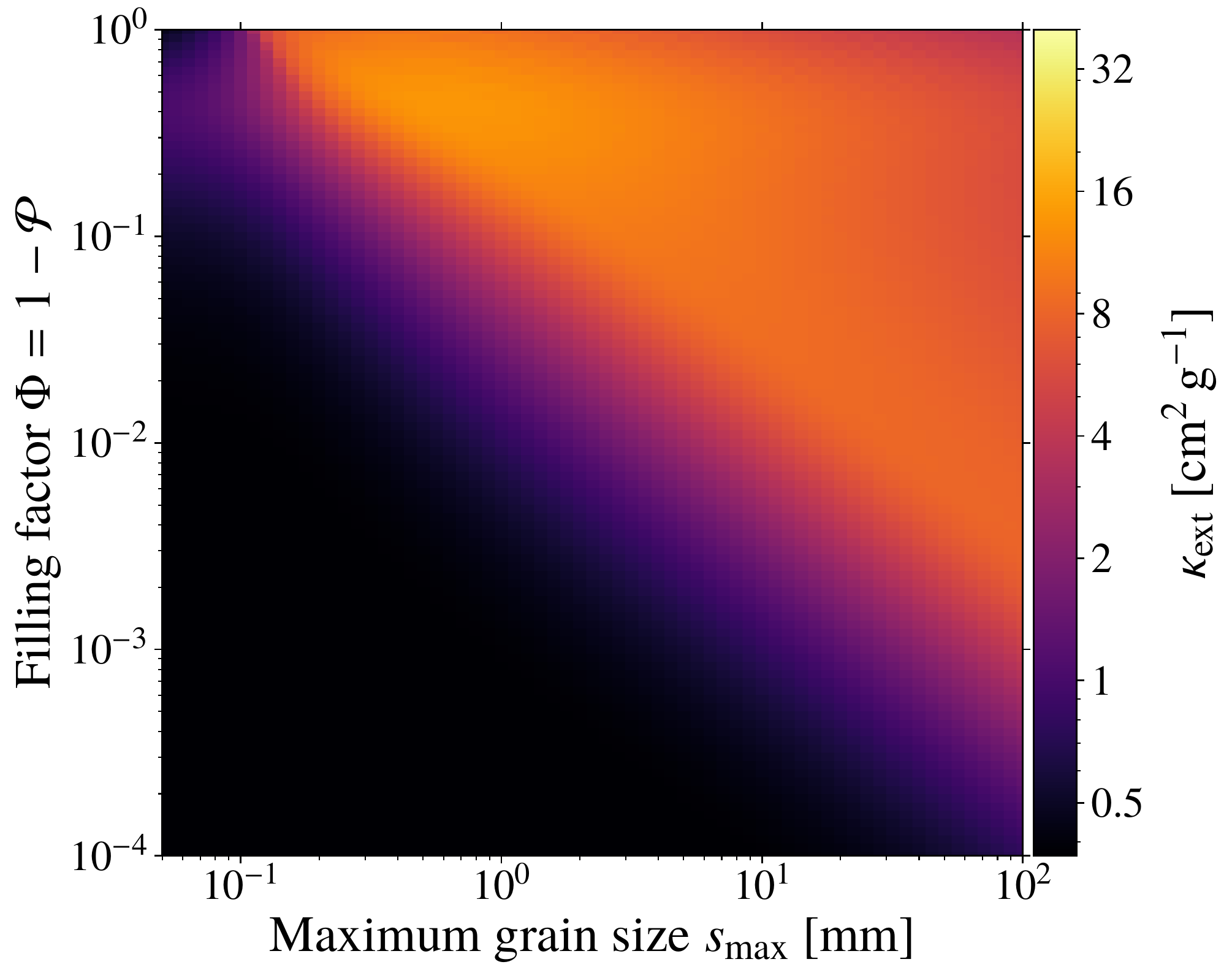}
            \includegraphics[width=0.49\linewidth, height=0.18\textheight]{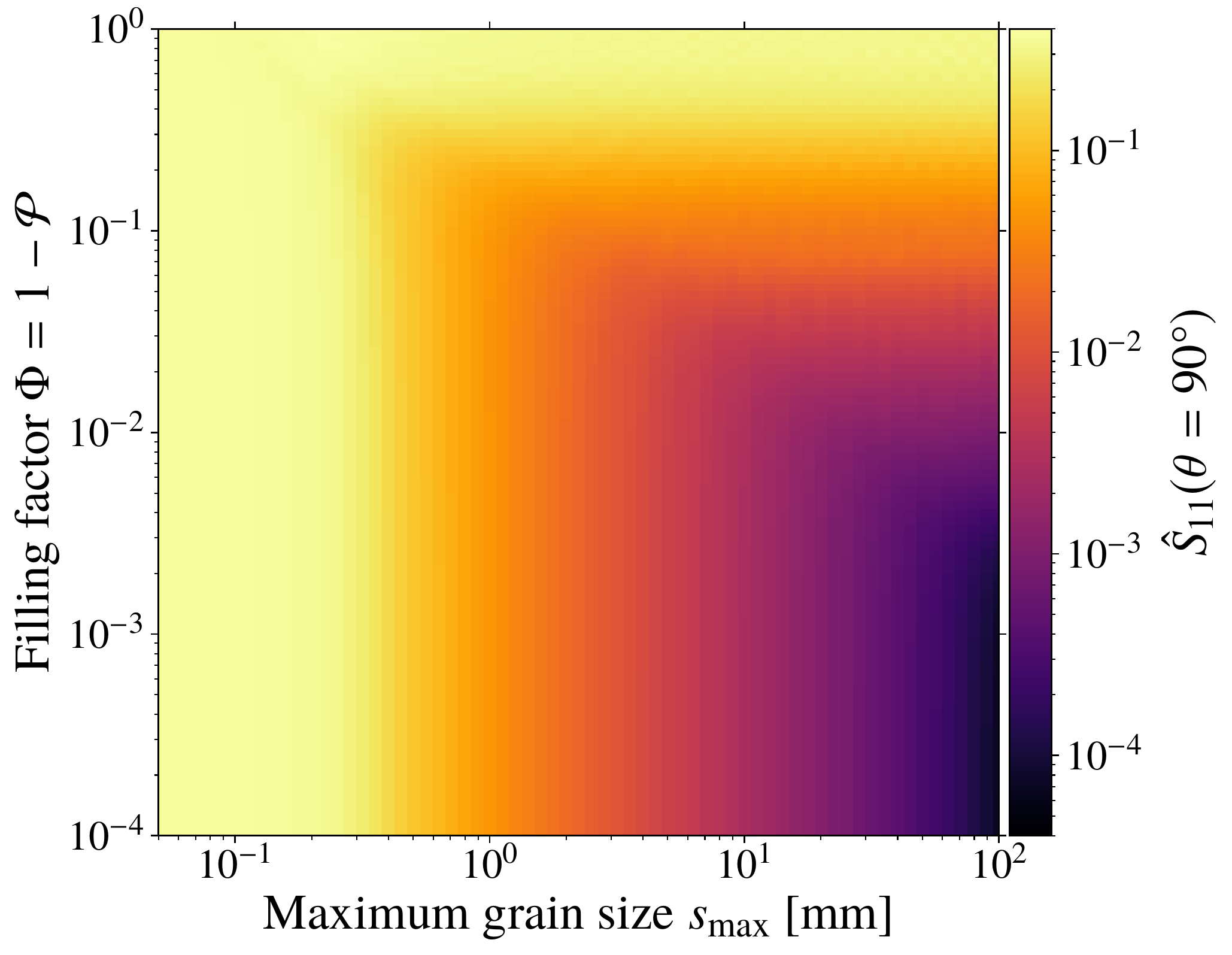}
            \caption{Degree of polarisation for single scattering $p = -\nicefrac{S_{12}}{S_{11}}$ for a scattering angle of $\theta=\SI{90}{\degree}$ \textit{(upper left)}, albedo \textit{(upper right)}, extinction opacity \textit{(lower left)}, and probability density function $\hat{S}_{11}$ for the scattering angle $\theta=\SI{90}{\degree}$ \textit{(lower right)}. All quantities were calculated for a grain size distribution with $q=\num{-3.8}$, $s_{\text{min}}=\SI{5}{\nm}$, and different upper grain size limits $s_{\text{max}}$ at a wavelength of \SI{850}{\um} for astrosil.}
            \label{fig:fill_smax_850_38}
        \end{figure*}
        \begin{figure*}
            \includegraphics[width=0.49\linewidth, height=0.18\textheight]{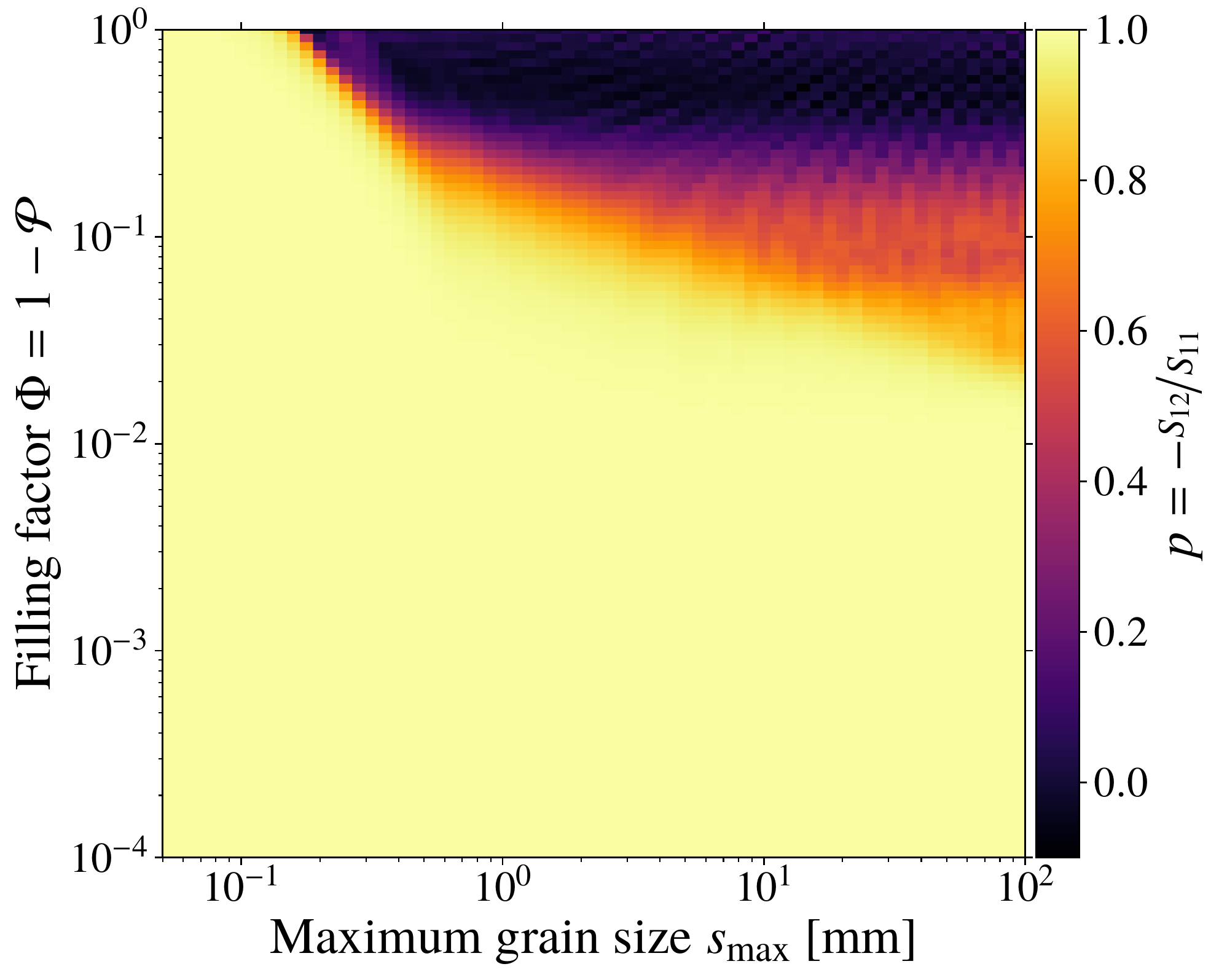}
            \includegraphics[width=0.49\linewidth, height=0.18\textheight]{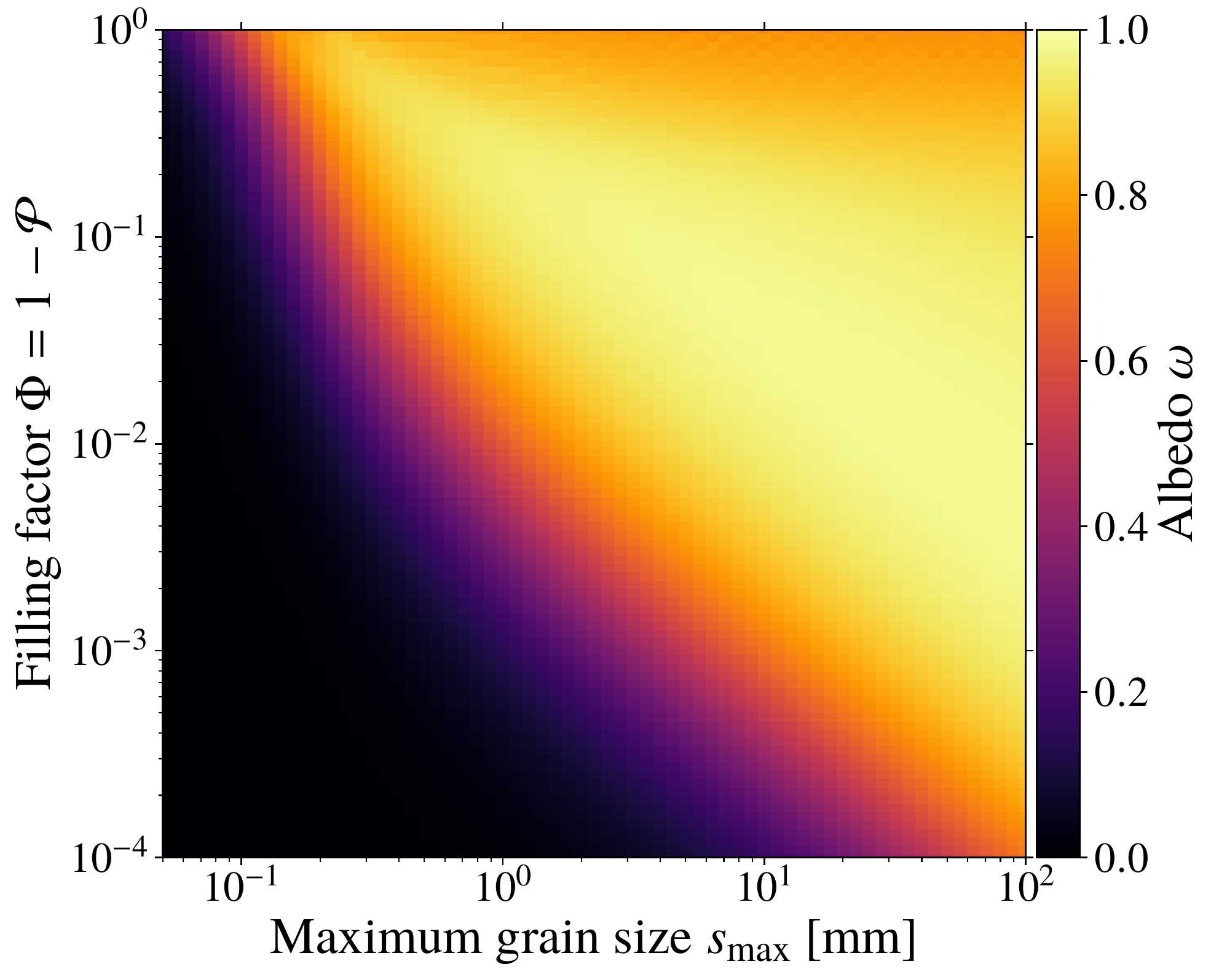}\\
            \includegraphics[width=0.49\linewidth, height=0.18\textheight]{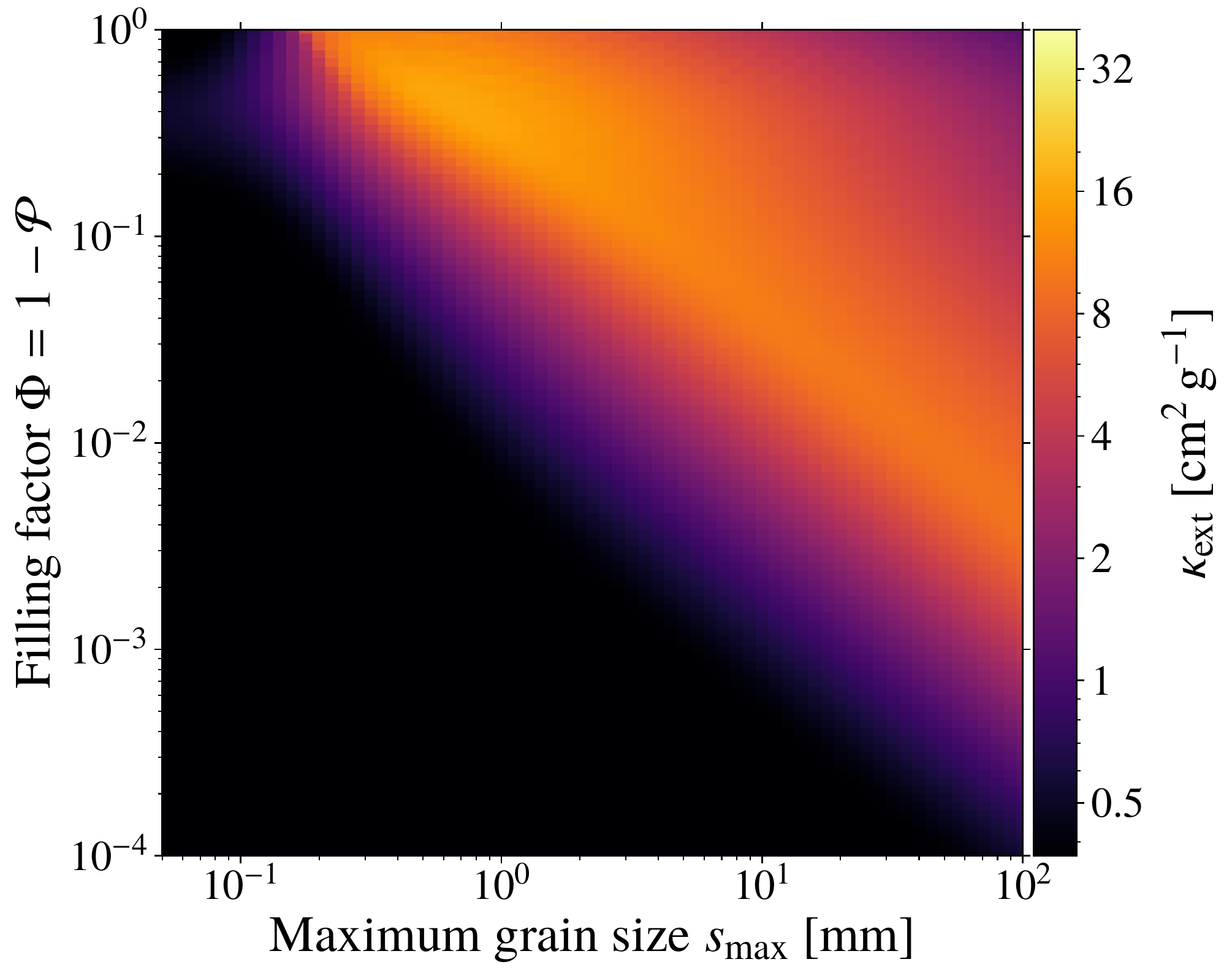}
            \includegraphics[width=0.49\linewidth, height=0.18\textheight]{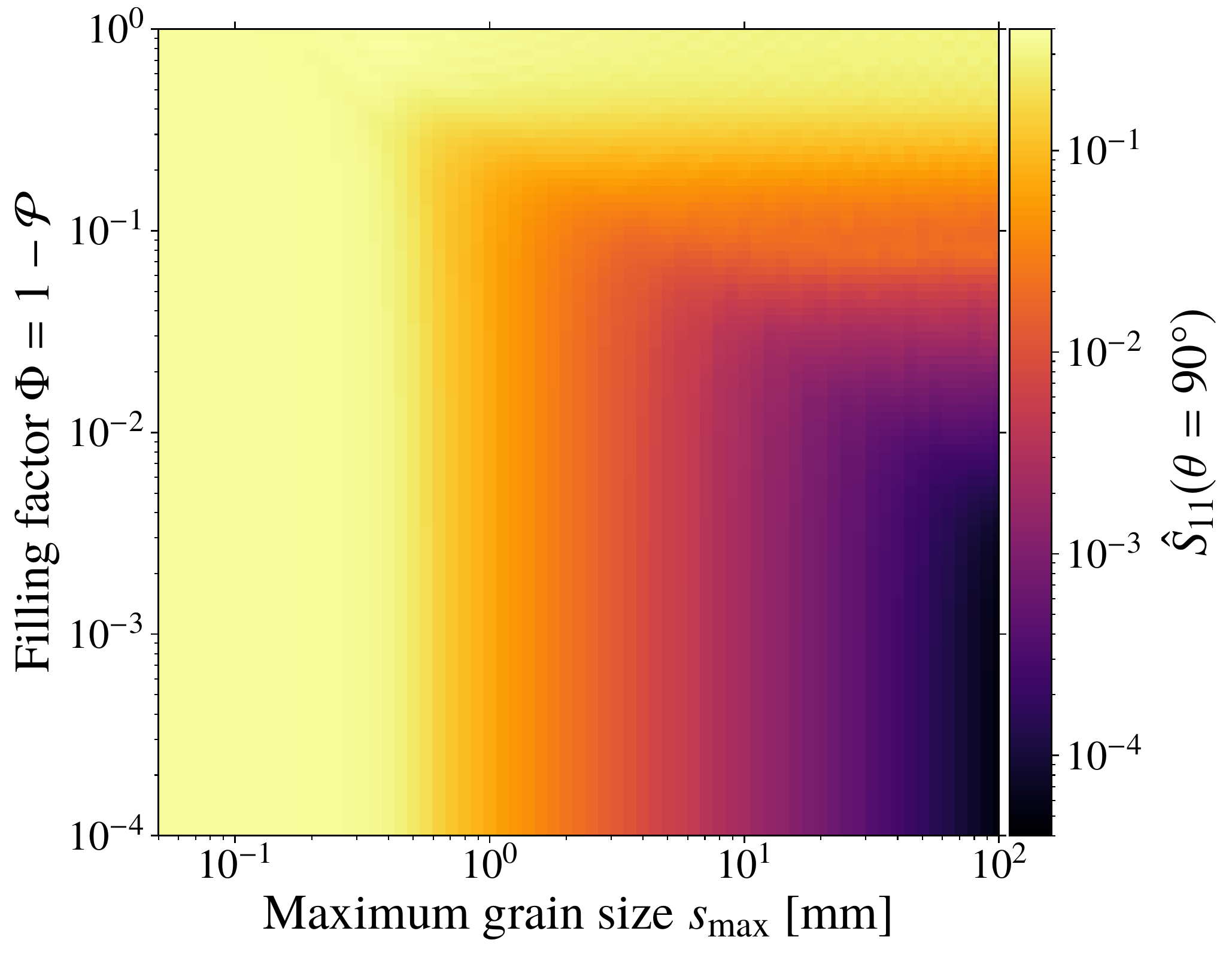}
            \caption{Degree of polarisation for single scattering $p = -\nicefrac{S_{12}}{S_{11}}$ for a scattering angle of $\theta=\SI{90}{\degree}$ \textit{(upper left)}, albedo \textit{(upper right)}, extinction opacity \textit{(lower left)}, and probability density function $\hat{S}_{11}$ for the scattering angle $\theta=\SI{90}{\degree}$ \textit{(lower right)}. All quantities were calculated for a grain size distribution with $q=\num{-3.5}$, $s_{\text{min}}=\SI{5}{\nm}$, and different upper grain size limits $s_{\text{max}}$ at a wavelength of \SI{1.3}{\mm} for astrosil.}
            \label{fig:fill_smax_1300_sili}
        \end{figure*}
        \begin{figure*}
            \includegraphics[width=0.49\linewidth, height=0.18\textheight]{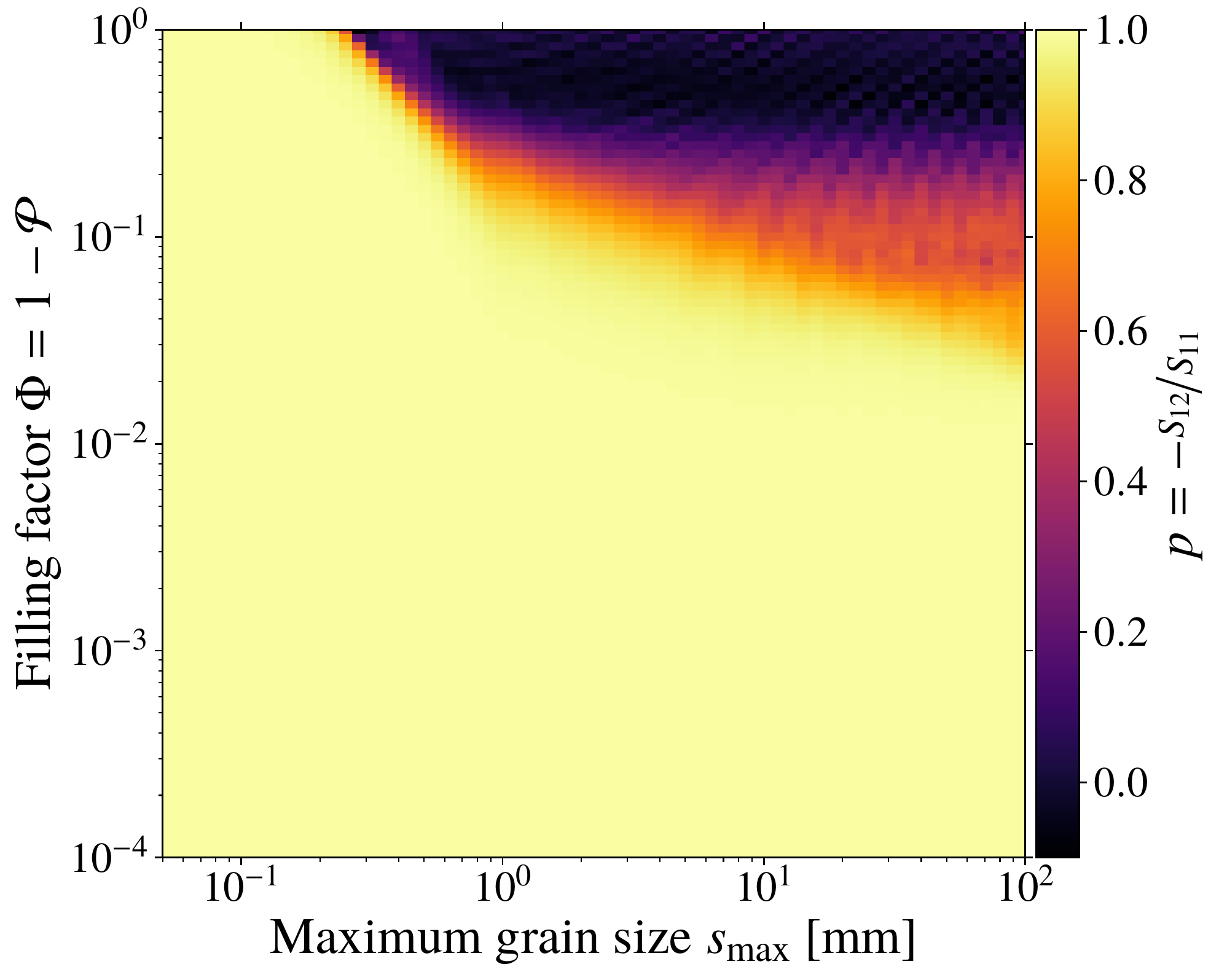}
            \includegraphics[width=0.49\linewidth, height=0.18\textheight]{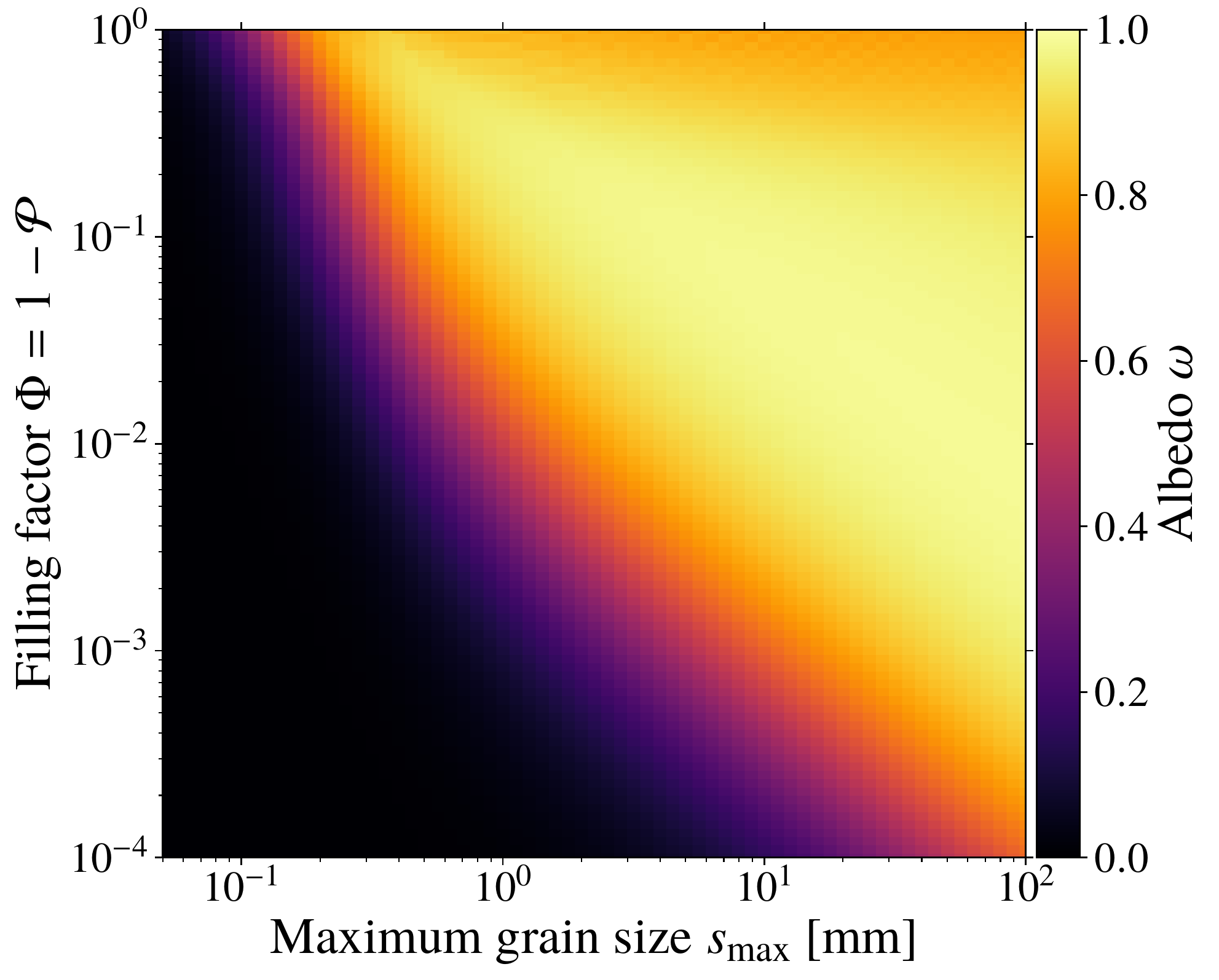}\\
            \includegraphics[width=0.49\linewidth, height=0.18\textheight]{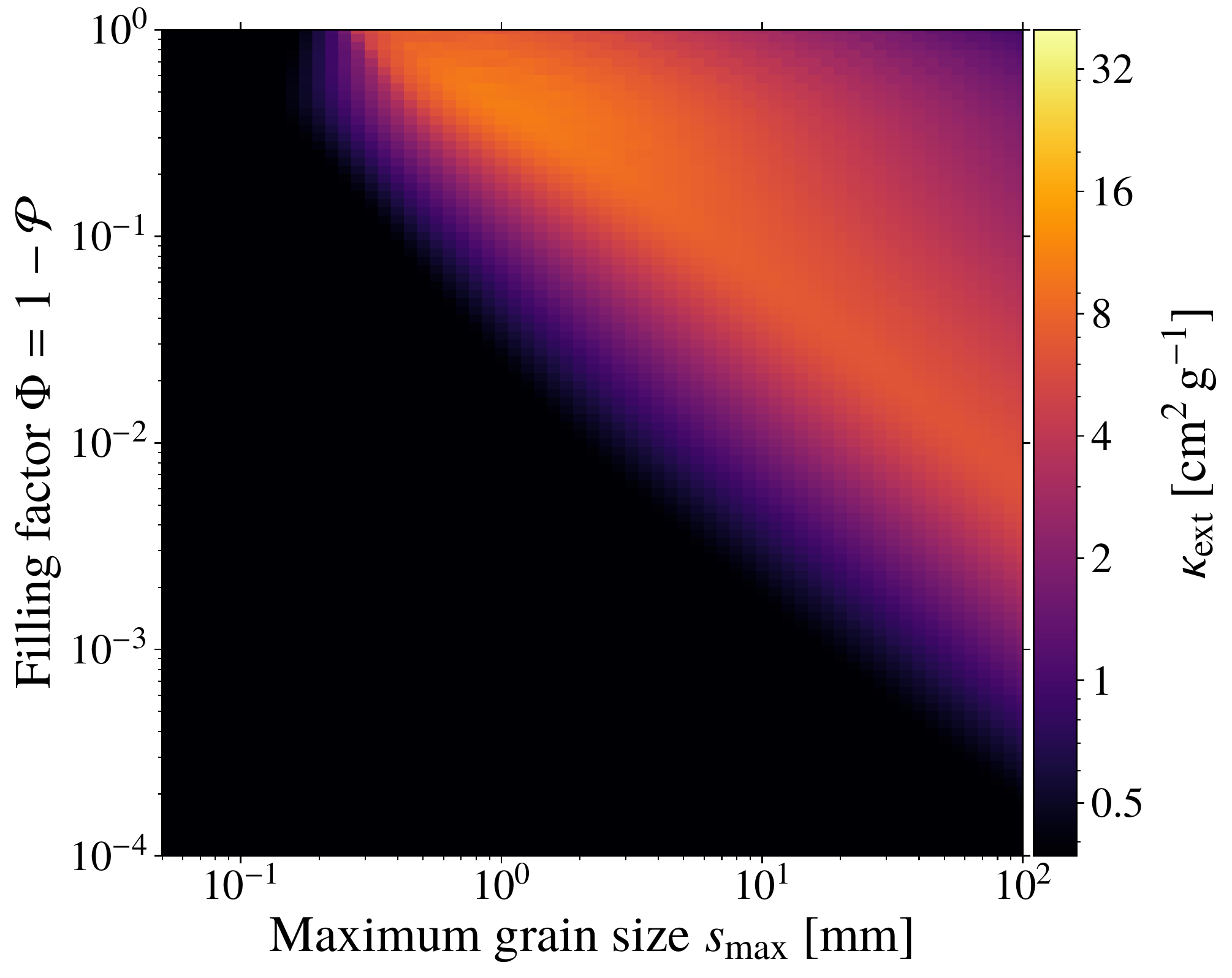}
            \includegraphics[width=0.49\linewidth, height=0.18\textheight]{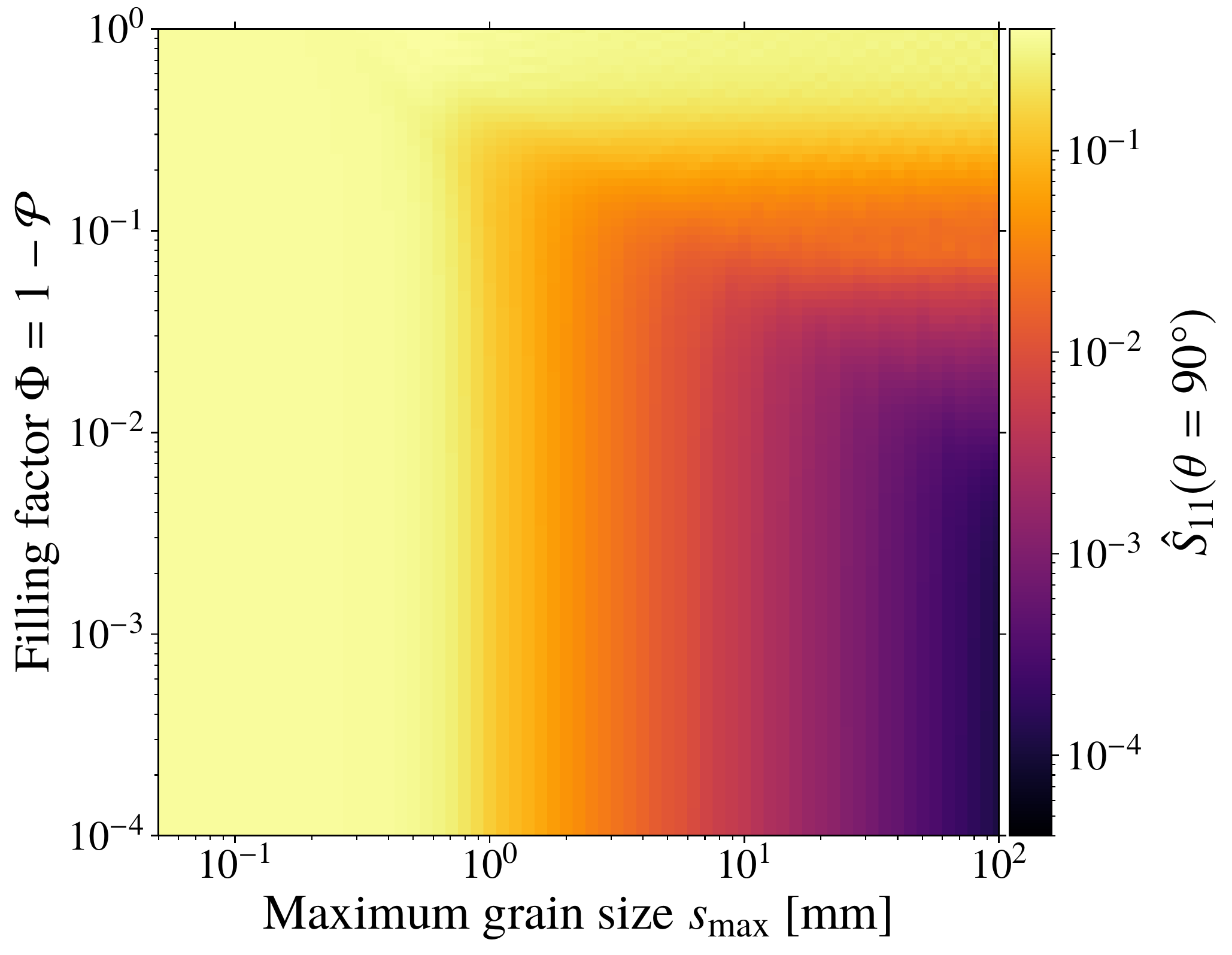}
            \caption{Degree of polarisation for single scattering $p = -\nicefrac{S_{12}}{S_{11}}$ for a scattering angle of $\theta=\SI{90}{\degree}$ \textit{(upper left)}, albedo \textit{(upper right)}, extinction opacity \textit{(lower left)}, and probability density function $\hat{S}_{11}$ for the scattering angle $\theta=\SI{90}{\degree}$ \textit{(lower right)}. All quantities were calculated for a grain size distribution with $q=\num{-3.5}$, $s_{\text{min}}=\SI{5}{\nm}$, and different upper grain size limits $s_{\text{max}}$ at a wavelength of \SI{2}{\mm} for astrosil.}
            \label{fig:fill_smax_2000_sili}
        \end{figure*}
\end{appendix}

\end{document}